 \documentclass[final,3p,times,twocolumn]{elsarticle}
\usepackage{amssymb}
\usepackage{amsmath}

\usepackage{graphicx}% Include figure files
\usepackage{dcolumn}% Align table columns on decimal point
\usepackage{bm}% bold math
\usepackage[usenames, dvipsnames]{xcolor} %ADDED
\usepackage{algorithm} %% ADDED
\usepackage{tikz} %% ADDED
\usepackage{setspace} %% ADDED

\usetikzlibrary{math} %%ADDED %needed tikz library
\usetikzlibrary{decorations.markings} % ADDED

\usepackage{hyperref}% add hypertext capabilities

\newcommand{\RedDottedCircle}{\begin{tikzpicture} 
\draw [dotted, color=red,line width=0.3mm] (-0.25,0) -- (0.25,0); 
\draw [color=red,line width=0.3mm] (0,0) circle (0.1);  
\end{tikzpicture}}

\newcommand{\BlueDottedTriangle}{\begin{tikzpicture} 
\draw [dotted, color=blue,line width=0.3mm] (-0.25,0) -- (0.25,0); 
\draw [color=blue,line width=0.3mm] (-0.1,-0.866*0.2/2.0) -- (0.1,-0.866*0.2/2.0) ;  
\draw [color=blue,line width=0.3mm] (-0.1,-0.866*0.2/2.0) -- (0.0,0.866*0.2/2.0) ;  
\draw [color=blue,line width=0.3mm] (0.0,0.866*0.2/2) -- (0.1,-0.866*0.2/2.0) ;  
\end{tikzpicture}}

\newcommand{\BlackDottedSquare}{\begin{tikzpicture} 
\draw [dotted, color=black,line width=0.3mm] (-0.25,0) -- (0.25,0); 
\draw [color=black,line width=0.3mm] (-0.1,-0.1) -- (0.1,-0.1) ;  
\draw [color=black,line width=0.3mm] (-0.1,-0.1) -- (-0.1,0.1) ;  
\draw [color=black,line width=0.3mm] (-0.1,0.1) -- (0.1,0.1) ; ;  
\draw [color=black,line width=0.3mm] (0.1,0.1) -- (0.1,-0.1) ; ;  
\end{tikzpicture}}

\newcommand{\GreenDottedSquare}{\begin{tikzpicture} 
\draw [dotted, color=OliveGreen,line width=0.3mm] (-0.25,0) -- (0.25,0); 
\draw [color=OliveGreen,line width=0.3mm] (-0.1,-0.1) -- (0.1,-0.1) ;  
\draw [color=OliveGreen,line width=0.3mm] (-0.1,-0.1) -- (-0.1,0.1) ;  
\draw [color=OliveGreen,line width=0.3mm] (-0.1,0.1) -- (0.1,0.1) ; ;  
\draw [color=OliveGreen,line width=0.3mm] (0.1,0.1) -- (0.1,-0.1) ; ;  
\end{tikzpicture}}

\newcommand{\BlackLine}{\begin{tikzpicture} 
\draw [color=white,line width=0.3mm] (0,0) circle (0.1);  
\draw [color=black,line width=0.3mm] (-0.25,0) -- (0.25,0); 
\end{tikzpicture}}

\newcommand{\RedCircle}{\begin{tikzpicture} 
\draw [color=white,line width=0.3mm] (-0.25,0) -- (0.25,0); 
\draw [color=red,line width=0.3mm] (0,0) circle (0.1);  
\end{tikzpicture}}

\newcommand{\BlueTriangle}{\begin{tikzpicture} 
\draw [color=white,line width=0.3mm] (-0.25,0) -- (0.25,0); 
\draw [color=blue,line width=0.3mm] (-0.1,-0.866*0.2/2.0) -- (0.1,-0.866*0.2/2.0) ;  
\draw [color=blue,line width=0.3mm] (-0.1,-0.866*0.2/2.0) -- (0.0,0.866*0.2/2.0) ;  
\draw [color=blue,line width=0.3mm] (0.0,0.866*0.2/2) -- (0.1,-0.866*0.2/2.0) ;  
\end{tikzpicture}}

\newcommand{\BlackSquare}{\begin{tikzpicture} 
\draw [color=white,line width=0.3mm] (-0.25,0) -- (0.25,0);
\draw [color=black,line width=0.3mm] (-0.1,-0.1) -- (0.1,-0.1) ;  
\draw [color=black,line width=0.3mm] (-0.1,-0.1) -- (-0.1,0.1) ;  
\draw [color=black,line width=0.3mm] (-0.1,0.1) -- (0.1,0.1) ; ;  
\draw [color=black,line width=0.3mm] (0.1,0.1) -- (0.1,-0.1) ; ;  
\end{tikzpicture}}

\newcommand{\GreenSquare}{\begin{tikzpicture} 
\draw [color=white,line width=0.3mm] (-0.25,0) -- (0.25,0);
\draw [color=OliveGreen,line width=0.3mm] (-0.1,-0.1) -- (0.1,-0.1) ;  
\draw [color=OliveGreen,line width=0.3mm] (-0.1,-0.1) -- (-0.1,0.1) ;  
\draw [color=OliveGreen,line width=0.3mm] (-0.1,0.1) -- (0.1,0.1) ; ;  
\draw [color=OliveGreen,line width=0.3mm] (0.1,0.1) -- (0.1,-0.1) ; ;  
\end{tikzpicture}}

\newcommand{\BlackCross}{\begin{tikzpicture} 
\draw [color=white,line width=0.3mm] (-0.25,0) -- (0.25,0);
\draw [color=black,line width=0.3mm] (-0.1,0) -- (0.1,0); 
\draw [color=black,line width=0.3mm] (0,-0.1) -- (0,0.1); 
\end{tikzpicture}}

\begin{document}

\begin{frontmatter}



\author[label1]{Florian Renard\corref{cor1}}
\ead{renard@cerfacs.fr}
 
\author[label2]{Yongliang Feng}
\ead{yongliang.feng@univ-amu.fr}

\author[label1]{Jean-Fran\c cois Boussuge}
\ead{boussuge@cerfacs.fr}

\author[label2]{Pierre Sagaut}
\ead{pierre.sagaut@univ-amu.fr}

\address[label1]{CERFACS, 42 Avenue G. Coriolis, 31057 Toulouse cedex, France}
\address[label2]{Aix Marseille Univ, CNRS, Centrale Marseille, M2P2, 13451 Marseille, France.}

\title{Improved compressible Hybrid Lattice Boltzmann Method \\ on standard lattice for subsonic and supersonic flows}


\author{}

\address{}

\begin{abstract}
A $D2Q9$ Hybrid Lattice Boltzmann Method (HLBM) is proposed for the simulation of both compressible subsonic and supersonic flows. This HLBM is an extension of the model of Feng $et$ $al.$~\cite{Feng2019}, which has been found, $via$ different test cases, to be unstable for supersonic regimes. 
The improvements consist of: (1) a new discretization of the lattice closure correction term making possible to properly simulate supersonic flows, (2) a corrected viscous stress tensor that takes into account polyatomic gases, and (3) a novel discretization of the viscous heat production term fitting with the regularized formalism.
The result is a hybrid method that resolves the mass and momentum equations with an LBM algorithm, and resolves the entropy-based energy equation with a finite volume method. This approach fully recovers the physics of the Navier-Stokes-Fourier equations with the ideal gas equation of state, and is valid from subsonic to supersonic regimes. It is then successfully assessed with both smooth flows and flows involving shocks. The proposed model is shown to be an efficient, accurate, and robust alternative to classic Navier-Stokes methods for the simulation of compressible flows.

\end{abstract}


\end{frontmatter}

%\linenumbers

\section{\label{sec:Intro}Introduction}

The accuracy, the efficiency and scalability of the Lattice Boltzmann Method (LBM) on standard lattices ($D1Q3$, $D2Q9$ and $D3Q19$/$Q27$)  has been widely demonstrated over the past decades, e.g.~\cite{guo2013lattice,Kruger_Book_2017}, for aeronautics applications involving turbulence and acoustics~\cite{sengissen2015simulations,khorrami2019toward}. Nevertheless, most existing standard LBM schemes relying on these lattices are restricted to isothermal and weakly compressible simulations, leading to a limitation of  their range of application, particularly in the field of aeronautics. In order to perform fully compressible computations while retaining the advantages of the standard LBM, two key issues must be resolved: the compressibility defect (Galilean invariance) of standard lattices~\cite{dellar2014lattice,shan2019central}, and the inabilty to correctly resolve the energy equation~\cite{alexander1993lattice,guo2013lattice}.
The straightforward extension of the LBM to compressible flows is achieved by increasing the number of discrete velocities of the lattice, leading to the commonly known multi-speed method~\cite{alexander1993lattice}. However, most of the methods relying on the multi-speed approach either use off-lattice~\cite{watari2003two} methods, which require space interpolations to recover populations at grid nodes, or use other numerical schemes~\cite{kataoka2004lattice} such as finite difference or finite volume methods, which are computationally expensive compared to the collide-and-stream based LBM. These methods are employed because the use of a multi-speed lattice with the standard collide-and-stream BGK algorithm leads to severe instabilities~\cite{siebert2008lattice}. Recent works by Frapolli $et$ $al.$~\cite{Frapolli2015,Frapolli2016,Frapolli2016a}, using a high order lattice and an entropic formulation of the collision operator, showed computations of  supersonic test cases such as the shock-vortex interaction or transonic airfoil. However, the computational and memory cost of this method is still an open question, as the two populations are discretized using 343 velocities for 3D problems, which enables the recovery of the thermal physics of interest (arbitrary heat capacity ratio and Prandtl number). Furthermore, the entropic collision model requires the resolution of a minimization problem at each grid point and for each time step, potentially increasing the cost of the method.

Another form of the LBM for the computation of compressible thermal flows is the double distribution function (DDF) method\cite{he1998novel}. It consists in using an additional set of distribution functions to model the energy.  As in the multi-speed case, most of the models that simulate thermal compressible flows at high Mach numbers do not use the collide-and-stream method, but rather more robust numerical schemes~\cite{guo2015discrete,xu2010unified,feng2016compressible}. However, recent publication of Saadat $et$ $al.$~\cite{saadat2019lattice}, who use a DDF model on a standard $D2Q9$ lattice along with appropriate corrective terms and a shifted lattice~\cite{frapolli2016lattice}, successfully simulated supersonic flows while remaining to some extent within the LBM framework. Although this model is based on a standard lattice, the CPU demand induced by the shifted lattice, as well as its ease of implementation for boundary conditions and mesh refinement, remain an open question.

A third method to recover the full compressible Navier-Stokes-Fourier equations is the Hybrid LBM method (HBLM). This approach shares traits with the DDF method in that it is also a two-equation based LBM, but in this case the temperature is computed using a finite difference or finite volume scheme instead of using an additional population of distribution functions. This approach has been widely used to model the hydrodynamics in the Boussinesq approximation where the density variation appears only in a forcing term~\cite{lallemand2003theory,feng2018regularized} and the temperature is otherwise considered to be a passive scalar. To model fully compressible flows, an ideal gas coupling between the two systems is mandatory. However, as mentioned by Lallemand $et$ $al.$~\cite{lallemand2003hybrid} the standard BGK (Bhatnagar-Gross-Krook) operator is largely responsible for numerical instabilities when using a thermal LBM model. They proposed to use a Multiple Relaxation Times (MRT) operator to overcome this issue. Nonetheless, the manifolds relaxation times present in this operator make the LBM model case-dependent. Recently, Feng $et$ $al.$~\cite{Feng2019} succeeded in performing stable high-subsonic computations using a new Hybrid Regularized Recursive (HRR) collision operator~\cite{Jacob2019} based on a single degree of freedom, which increased the stability of the original recursive regularized approach~\cite{malaspinas2015increasing,Coreixas2017}. In their model, they use a standard $D2Q9$ lattice where the Galilean invariance defect is corrected using a Hermite-based force term computed with finite differences. The present work follows the path paved by this new approach, and further extends it to the simulation of transonic and supersonic flows involving discontinuities.

The paper is articulated in five parts. In Sec.~\ref{sec:Model overview} the general construction and associated macroscopic equations of the hybrid scheme of~\cite{Feng2019} is introduced. In Sec.~\ref{sec: Improvements}, the space and time discretization of the scheme is derived with its threefold improvements: (1) a new corrective term that correctly retrieves the viscous stress tensor, (2) a novel discretization technique of the viscous heat production term in the energy equation and (3) a tailored numerical scheme for the compressibility correction terms that increases stability for high Mach number flows. 
In Sec.~\ref{sec: Validation of the improvements}, the improvements of the model are validated through several canonical test cases assessing: (1) the viscous dissipation/production and thermal dissipation, (2) the capacity of the scheme to capture vortical flows and acoustics in subsonic and supersonic regimes. Throughout this validation stage, the improved scheme is compared to the original one, which is shown to be restricted to the computation of subsonic flows due to stability problems.
In Sec.~\ref{sec:Numerical validation}, the acoustic validation of the improved model is completed. This is followed by two test cases aimed at assessing: (1) the ability to handle flow with discontinuities (2) the ability to capture non-linear acoustics in a transonic case. Finally, conclusions are drawn in Sec.\ref{sec:Conclusion}.

\section{\label{sec:Model overview} Hybrid Discrete Velocity Boltzmann Equation on standard lattice}

In this section, the continuous equations in space and time resolved by the hybrid model~\cite{Feng2019} are presented. First, the governing equations are given in their general form. Next, the choice of the equilibrium distribution function is detailed. The thermal and compressible restrictions of the standard lattices are then presented, and the correction term allowing to solve the compressibility issue is introduced. Afterwards, the Chapmann-Enskog expansion is used to retrieve the equivalent macroscopic equations solved by the hybrid model. Finally, the speed of sound relation generated by the thermodynamic closure of the hybrid model will be discussed.

\subsection{\label{subsec: Gov equations} Governing equations}

In this Hybrid model, the mass and momentum are computed by the Discrete Velocity Boltzmann Equation (DVBE) $\mathcal{B}_i$, which determines the spatiotemporal evolution of the discrete particle distribution function $f_i$. This equation reads:
\begin{equation}
\mathcal{B}_i : \frac{\partial f_i}{\partial t} + c_{i,\alpha}\frac{\partial f_i}{\partial x_\alpha} = -\frac{1}{\tau}\left( f_i - f_i^{eq} \right) + \psi_i,\quad \forall i \in \left[0,m-1\right],
\label{eq: DVBE eq}
\end{equation}
where the BGK collision operator~\cite{bhatnagar1954model} has been used. $\tau$ is the characteristic time for the relaxation of the distribution function $f_i$ toward $f_i^{eq}$, the local thermodynamic equilibrium distribution function, detailed in Sec.~\ref{sec:Equlilibrium}. $\psi_i$ refers to a correction term defined in Sec.~\ref{sec:closure relation and corr term} and $c_{i,\alpha}$ to the discrete microscopic velocities where the Greek subscripts $\alpha$ denotes the spatial directions in Cartesian coordinates. These velocities span the so-called lattice $DnQm$, with $n$ the spatial dimension and $m$ the number of discrete velocities. A Gauss-Hermite quadrature~\cite{shan2006kinetic} ensuring a fifth-order recovery of the weighted Hermite polynomial yields the $D2Q9$ lattice~\cite{qian1992lattice}, whose velocities are defined as:
\begin{align}
c_{i,x} = [0,1,0,-1,0,1,-1,-1,1]  C_0, \\
c_{i,y} = [0,0,1,0,-1,1,1,-1,-1]  C_0, \\
\nonumber
\end{align} 
where $C_0=\sqrt{rT_{r}}/c_s$ with $T_r$ a reference temperature, $r$ the molecular gas constant (taken at $r=287.15 \,\mathrm{Km}^2\mathrm{s}^{-2}$ in the following), and $c_s=1/\sqrt{3}$ the so-called lattice constant.
The associated lattice weights read:
\begin{equation}
w_i = \left[\frac{4}{9},\frac{1}{9},\frac{1}{9},\frac{1}{9},\frac{1}{9},\frac{1}{36},\frac{1}{36},\frac{1}{36},\frac{1}{36} \right]. 
\end{equation} 
In the following, all variables are considered dimensionless using $T_r$, $\rho_r$ as a reference density, and $\Delta x$ and $\Delta t = \Delta x /C_0$ being respectively an arbitrary length and time.
Computing the moments of the distribution functions yields the mass at the zeroth order:
\begin{equation}
\rho \equiv \sum_{i=0}^{m-1}f_i=\sum_{i=0}^{m-1}f_i^{eq},
\label{eq: zero moment of f}
\end{equation} 
and the momentum at first-order:
\begin{equation}
\rho u_\alpha\equiv \sum_{i=0}^{m-1} c_{i,\alpha} f_i = \sum_{i=0}^{m-1} c_{i,\alpha} f_i^{eq},
\label{eq: firts moment of f}
\end{equation} 
which are collision invariant by construction.

The Hybrid approach uses an entropy equation to compute the dimensionless temperature $\theta=T/T_r$ of the system. The usual equation for entropy is given by:
\begin{equation}
\frac{\partial s}{ \partial t} + u_\alpha \frac{\partial s}{ \partial x_\alpha}   = - \frac{1}{ \rho \theta}\frac{\partial }{ \partial x_\alpha} \left(- \lambda  \frac{\partial \theta}{ \partial x_\alpha}  \right) +   \frac{\Phi}{ \rho \theta}, 
\label{eq: Entropy eq}
\end{equation}
where $\lambda=\mu / \left(c_p\mathrm{Pr}\right)$ is the heat conductivity and $\Phi$ the viscous heat production term defined as:
\begin{equation}
\Phi = \mu\left( \frac{\partial u_\alpha}{\partial x_\beta} + \frac{\partial u_\beta}{\partial x_\alpha} - \frac{2}{D} \frac{\partial u_\gamma}{\partial x_\gamma} \delta_{\alpha\beta} \right) \frac{\partial u_\alpha}{\partial x_\beta}.
\label{eq:viscous heat}
\end{equation}
$\delta_{\alpha\beta}$ refers to the kronecker symbol and $\mathrm{Pr}$, $\mu$ and $D$ are respectively the Prandtl number, the dynamic viscosity and the spatial dimension of the system. The entropy $s$ is directly derived from the Gibbs equation and reads:
\begin{equation}
\label{eq: Entropie expression}
s= c_v\ln \left(\frac{\theta}{\rho ^{(\gamma_g -1)}} \right),
\end{equation}
where $\gamma_g=c_p/c_v$ refers to the heat capacity ratio with $c_v$ and $c_p$ respectively the specific heat at constant volume and pressure. 

Finally, the Hybrid DVBE system on the standard $D2Q9$ lattice consists of a set of ten partial differential equations (PDE). Eq.~(\ref{eq: DVBE eq}) resolves the mass and the momentum while a single PDE Eq.~(\ref{eq: Entropy eq}), is used to solve the entropy from which the temperature is obtained. The dependency of the entropy equation on the DVBE can be explicitly seen through $\rho$ and $u_\alpha$, while the thermal coupling of the entropy equation to the DVBE is described in Sec.~\ref{sec:Equlilibrium}.

\subsection{\label{sec:Equlilibrium}The equilibrium distribution function}

The usual expansion of the equilibrium distribution function $f^{eq}$ on Hermite polynomials~\cite{Shan1998} is adopted  and reads for the discrete velocity case: 
\begin{equation}
f_i^{eq} = w_i \sum_{n=0}^{N} \frac{1}{n!c_s^{2n}}\boldsymbol{a}^{eq,\left(n\right)} : \boldsymbol{\mathcal{H}}_i^{\left(n\right)},
\label{eq:Equilibrium}
\end{equation}
where $:$ stands for the full contraction of indexes, also known as Frobenius inner product. The coefficients $\boldsymbol{a}^{eq,\left(n\right)}$ are the $n^{th}$-order Hermite moments of the Maxwell-Boltzmann equilibrium distribution function. Up to the fourth order, they are given by:
\begin{subequations}
\begin{align}
\label{eq: a0}&a^{eq}_{0}= \rho,  \\ 
\label{eq: a1}&a^{eq}_{\alpha}= \rho u_\alpha, \\
\label{eq: a2}&a^{eq}_{\alpha\beta}= \rho u_\alpha u_\beta + \rho c_s^2 \left( \theta-1 \right) \delta_{\alpha\beta},  \\
\label{eq: a3}& a^{eq}_{\alpha\beta\gamma}= \rho u_\alpha u_\beta u_\gamma \nonumber \\
&\quad\quad\quad+ \rho c_s^2 \left( \theta-1 \right) \left( u_\alpha \delta_{\beta\gamma} + u_\beta \delta_{\alpha\gamma} + u_\gamma \delta_{\alpha\beta} \right),  \\
\label{eq: a4}&a^{eq}_{\alpha\beta\gamma\delta} = \rho u_\alpha u_\beta u_\gamma u_\delta \nonumber \\
&\quad\quad\quad+ \rho c_s^4 \left( \theta-1 \right)^2 \left( \delta_{\alpha\beta}\delta_{\gamma\delta} + \delta_{\alpha\gamma}\delta_{\beta\delta} + \delta_{\alpha\delta}\delta_{\beta\gamma}   \right) \nonumber  \\
&\quad\quad\quad+ \rho c_s^2 \left( \theta-1 \right) \left( u_\alpha u_\beta \delta_{\gamma\delta} + u_\alpha u_\gamma \delta_{\beta\delta} + u_\alpha u_\delta \delta_{\beta\gamma} \right. \nonumber\\ 
&\left. \quad\quad\quad+ u_\beta u_\gamma \delta_{\alpha\delta} + u_\beta u_\delta \delta_{\alpha\gamma} + u_\gamma u_\delta \delta_{\alpha\beta} \right).   
\end{align}
\end{subequations}
$\boldsymbol{\mathcal{H}}_{i}$ refers to the discrete Hermite polynomial in terms of velocity, defined up to the fourth order as:
\begin{subequations}
\begin{align}
\label{eq: H0}&\mathcal{H}_{i,0}=1, \\
\label{eq: H1}&\mathcal{H}_{i,\alpha}= c_{i,\alpha},  \\
\label{eq: H2}&\mathcal{H}_{i,\alpha\beta}= c_{i,\alpha}c_{i,\beta} - c_s^2 \delta_{\alpha\beta},  \\
\label{eq: H3}&\mathcal{H}_{i,\alpha\beta\gamma}=c_{i,\alpha}c_{i,\beta}c_{i,\gamma} - c_s^2 \left( c_{i,\alpha}\delta_{\beta\gamma} + c_{i,\beta}\delta_{\alpha\gamma} \right. \nonumber \\
&\qquad\qquad \left. + c_{i,\gamma}\delta_{\alpha\beta} \right),  \\
\label{eq: H4}&\mathcal{H}_{i,\alpha\beta\gamma\delta}=c_{i,\alpha}c_{i,\beta}c_{i,\gamma}c_{i,\delta} - c_s^2\left( c_{i,\alpha}c_{i,\beta}\delta_{\gamma\delta} + c_{i,\alpha}c_{i,\gamma}\delta_{\beta\delta} \right. \nonumber \\ 
&\qquad\qquad \left.+ c_{i,\alpha}c_{i,\delta}\delta_{\beta\gamma} + c_{i,\beta}c_{i,\gamma}\delta_{\alpha\delta} + c_{i,\beta}c_{i,\delta}\delta_{\alpha\gamma} \right. \nonumber \\
&\qquad\qquad \left. + c_{i,\gamma}c_{i,\delta}\delta_{\alpha\beta}\right) \nonumber \\
&\qquad\qquad + c_s^4 \left( \delta_{\alpha\beta}\delta_{\gamma\delta} + \delta_{\alpha\gamma}\delta_{\beta\delta} + \delta_{\alpha\delta}\delta_{\beta\gamma} \right).   
\end{align}
\end{subequations}
Unlike the standard isothermal LBM, the present equilibrium distribution function is developed in term of temperature giving rise to a perfect gas thermodynamic closure $p=\rho c_s^2 \theta$. This relation, shown later in Sec.~\ref{sec:Macro equation}, appears naturally once the macroscopic equation are recovered using the Chapman-Enskog development. In the following, only the moments computable with the $D2Q9$ basis described in the next section~\ref{sec:closure relation and corr term} are kept to build the equilibrium distribution function.

\subsection{\label{sec:closure relation and corr term}Lattice closure relation and correction term}

For the $D2Q9$ lattice, the discrete equilibrium distribution functions $f^{eq}_i$ truncated at the second-order in terms of Hermite series expansion, allows for recovering the zeroth, first, and second-order equilibrium moments without error. However, higher order moments are not properly recovered due to the closure relation~\cite{Karlin2010}. This relation, also referred to the low symmetry of the lattice~\cite{Li2012}, is the root cause limitation of the lattice to recover the correct moments of the equilibrium distribution function. This gives rise to a linear dependency of higher order moments on lower order ones. For the $D2Q9$ lattice this relation reads:
\begin{equation}
c_{i,\alpha}^3=c_{i,\alpha},
\end{equation}
which can be rewritten  in terms of Hermite polynomial as:
\begin{equation}
\mathcal{H}_{i,\alpha\alpha\alpha}= \left(1-3c_s^2 \right) \mathcal{H}_{i,\alpha}=0.
\end{equation}
This limitation leads to a biased evaluation of the third and higher order moments which reduces the ability of the lattice to recover the desired set of macroscopic equations. The third-order equilibrium moment is therefore null: 
\begin{equation}
a^{eq}_{\alpha\alpha\alpha} = \sum_{i=0}^{m-1} \mathcal{H}_{i,\alpha\alpha\alpha} f_i^{eq} = 0.
\label{eq: closure a3 Q9 Q27}
\end{equation}
Once a Chapman-Enskog expansion is carried out (see Sec.~\ref{sec:Macro equation}), it is clear that this closure relation is directly responsible for the well-known Galilean invariance issue observed on  standard lattices ($D1Q3$, $D2Q9$, $D3Q27$). Moreover, higher order moments associated to the temporal evolution of the energy are also biased leading  to the isothermal restriction. 		
The Hermite polynomials supported by the $D2Q9$ basis read~\cite{Mattila2017,Coreixas2017}:
\begin{equation}
\begin{split}
\mathcal{B}^{\mathcal{H}}_{D2Q9} = \left( \mathcal{H}_{i,0}, \mathcal{H}_{i,x}, \mathcal{H}_{i,y}, \mathcal{H}_{i,xx}, \mathcal{H}_{i,xy}, \mathcal{H}_{i,yy}, \right.\\
\left. \mathcal{H}_{i,xxy}, \mathcal{H}_{i,yyx}, \mathcal{H}_{i,xxyy} \right),
\end{split}
\end{equation}
with their associated Hermite equilibrium moments:
\begin{equation}
 \left( a^{eq}_{0}, a^{eq}_{x}, a^{eq}_{y}, a^{eq}_{xx}, a^{eq}_{xy}, a^{eq}_{yy},a^{eq}_{xxy}, a^{eq}_{yyx}, a^{eq}_{xxyy} \right),
\end{equation}
which are used to build the equilibrium distribution function Eq.~(\ref{eq:Equilibrium}). Developing $f_i^{eq}$ up to the fourth order for the present lattice increases the numerical numerical stability~\cite{wissocq2019extended,wissocq2019thesis}, which presents a further improvement over the original model, which goes only to third order~\cite{Feng2019}.

Thus, this closure relation results in a null third-order Hermite equilibrium moment $a^{eq}_{\alpha\alpha\alpha}$, whose spatial derivative is of paramount importance to recover the correct viscous part of the momentum equation as shown later in Sec.~\ref{sec:Macro equation}. However, this defect can be corrected by injecting a corrective term into the DVBE under the following form:
\begin{equation}
E_{1,\alpha\beta}= \frac{\partial}{\partial x_\alpha} \underbrace{\left( \rho u_\alpha\left(1-\theta -u_\alpha^2 \right) \right)}_{=-a^{eq}_{\alpha\alpha\alpha}}\delta_{\alpha\beta},
\end{equation}
which must appears in the viscous stress tensor of the momentum equation.
This is done through the force term $\psi_i$ in Eq.~(\ref{eq: DVBE eq}) which reads:
\begin{equation}
\begin{split}
\psi_i =& w_i\frac{\mathcal{H}_{i,\alpha\beta}}{2c_s^4} E_{1,\alpha\beta},
\end{split}
\end{equation}
where $E_{1,\alpha\beta}$ corresponds to the second-order Hermite moment of $\psi_i$.
In the standard isothermal case ($\theta=1$), this error only reduces to a cubic error term which can be neglected for low Mach number cases. However, for thermal cases where $\theta\neq1$, a first-order error term in velocity arises, which is of utmost importance and cannot be neglected anymore.
The corrective term includes a factor of the second-order Hermite polynomial, therefore it does not affect the computation of the macroscopic variables $\rho$ in Eq.~(\ref{eq: zero moment of f})  and $\rho u_\alpha$ in Eq.~(\ref{eq: firts moment of f}), thanks to the orthogonality properties of the Hermite polynomials~\cite{shan2006kinetic}. Thus, due to its second order nature, this term acts only on the viscous part of the momentum equation as shown in the next section.

\subsection{\label{sec:Macro equation}Thermo-hydrodynamic limits of the model}

The equivalent macroscopic equations associated to the DVBE (Eq.~(\ref{eq: DVBE eq})) are found by computing its velocity moments in the canonical basis after expanding the distribution functions in terms of the small parameter $\tau$, which is related to the Knudsen number. This expansion is the so-called Chapman-Enskog development~\cite{chapman1970mathematical} and reads at the first order:
\begin{equation}
f_i \simeq f_i^{eq}   - \tau \left( -\psi_i + \frac{\partial f_i^{eq}}{\partial t} + c_{i,\alpha}\frac{\partial f_i^{eq}}{\partial x_\alpha}    \right),
\label{eq:ChapmanO1 sec}
\end{equation}
or more commonly:
\begin{equation}
f_i \simeq f_i^{eq}   + f_i^{1},
\end{equation}
where $f_i^{1}$ corresponds to the first-order approximation in $\tau$ of the off-equilibrium part of $f_i$. More details on this development can be found in App.~\ref{sec:Chapman dev}.
Before computing the moments of Eq.~(\ref{eq: DVBE eq}), it is worth noting that its two first Hermite moments, computed respectively with Eq.~(\ref{eq: H0}) and Eq.~(\ref{eq: H1}), are strictly equivalent to the raw moments. For higher order moments, the recursive relation on the second-order Hermite polynomial gives:
\begin{equation}
\mathcal{H}_{i,\alpha}c_{i,\beta} = \mathcal{H}_{i,\alpha\beta} + c_s^2\delta_{\alpha\beta}\mathcal{H}_{i,0},
\label{eq:Hermite trick second}
\end{equation}
and similarly for the third-order polynomial:
\begin{equation}
\mathcal{H}_{i,\alpha\beta}c_{i,\gamma} = \mathcal{H}_{i,\alpha\beta\gamma} + c_s^2 \left( \mathcal{H}_{i,\alpha} \delta_{\beta\gamma} + \mathcal{H}_{i,\beta} \delta_{\alpha\gamma} \right),
\label{eq:Hermite trick third}
\end{equation}
will be useful for the following.
Computing now the zeroth and first-order moment of Eq.~(\ref{eq: DVBE eq}), with the use of Eq.~(\ref{eq: zero moment of f}) and Eq.~(\ref{eq: firts moment of f}), leads respectively to the mass equation:
\begin{equation}
\sum_{i=0}^{m-1}\mathcal{H}_{i,0}\mathcal{B}_i \rightarrow \frac{\partial \rho}{\partial t} + \frac{\partial \rho u_\alpha}{\partial x_\alpha} = 0,
\label{eq:mass eq}
\end{equation} 
and the momentum equation:
\begin{equation}
\sum_{i=0}^{m-1}\mathcal{H}_{i,\alpha}\mathcal{B}_i \rightarrow \frac{\partial \rho u_\alpha}{\partial t} + \frac{\partial \Pi_{\alpha\beta}}{\partial x_\beta} = 0.
\label{eq:momentum eq}
\end{equation} 
The second-order moment, thank to the use of Eq.~(\ref{eq:Hermite trick second}) is expressed as follows:
\begin{equation}
\Pi_{\alpha\beta} = \sum_{i=0}^{m-1} \mathcal{H}_{i,\alpha\beta}f_i + \sum_{i=0}^{m-1} c_s^2 \delta_{\alpha\beta}\mathcal{H}_{i,0}f_i.
\label{eq: Pi starting point}
\end{equation}
Substituting Eq.~(\ref{eq:ChapmanO1 sec}) in Eq.~(\ref{eq: Pi starting point}) , using Eq.~(\ref{eq:Hermite trick third}) and noting that the zeroth order off-equilibrium Hermite moment cancels out due to Eq.~(\ref{eq:mass eq}), this gives:
\begin{equation}
\begin{split}
\Pi_{\alpha\beta} =& \rho u_\alpha u_\beta + \rho c_s^2 \theta \delta_{\alpha\beta} - \tau \left( -E_{1,\alpha\beta} + \frac{\partial a^{eq}_{\alpha\beta}}{\partial t}   \right. \\
 & \left.+ \frac{\partial a^{eq}_{\alpha\beta\gamma}}{\partial x_\gamma}  + \frac{\partial}{\partial x_\gamma} \left[ \rho c_s^2 \left( u_{\alpha} \delta_{\beta\gamma} + u_{\beta} \delta_{\alpha\gamma} \right) \right] \right)   \\
 =&\Pi_{\alpha\beta}^{eq} + a^1_{\alpha\beta}. 
\end{split}
\label{eq: Pi inter}
\end{equation}
Here, $a^1_{\alpha\beta}$ is the second-order off-equilibrium Hermite moment and $E_{1,\alpha\beta}$ is the integrated correction term over the second-order Hermite polynomial Eq.~(\ref{eq: H2}).
Classically, Eq.~(\ref{eq:mass eq}) and Eq.~(\ref{eq:momentum eq}) yield the Euler equations when approximating each term by its equilibrium part. This is equivalent to keep only the terms in $\tau^0$ and results in: 
\begin{subequations}
\begin{align}
\label{eq: Mass from Euler}&\frac{\partial \rho}{\partial t} + \frac{\partial \rho u_\alpha}{\partial x_\alpha} = 0 \\
\label{eq: QDM from Euler}&\frac{\partial \rho u_\alpha}{\partial t} + \frac{\partial }{\partial x_\beta} \left(\rho u_\alpha u_\beta + p \delta_{\alpha\beta} \right) = 0,
\end{align}
\label{eq: Euler system}
\end{subequations}
with $p$ the pressure found to be :
\begin{equation}
p=\rho c_s^2 \theta.
\label{eq: perfect gas}
\end{equation}

Up to this point, one can clearly guess that the viscous contribution in the momentum equation will be given by $a^1_{\alpha\beta}$. Moreover, it is important to recall that the Chapman-Enskog expansion has been done up to the first-order in $\tau$ (or Knudsen). Thus no higher power contributions are retained to compute:
\begin{equation}
\frac{\partial a^{eq}_{\alpha\beta}}{\partial t} = \frac{\partial \rho u_\alpha u_\beta}{\partial t} + \frac{\partial p}{\partial t} \delta_{\alpha\beta} - \frac{\partial \rho c_s^2}{\partial t} \delta_{\alpha\beta},
\label{eq: tempo der a2eq}
\end{equation}
reducing the choice to Eq.~(\ref{eq: QDM from Euler}) for the computation of the quadratic term in the temporal derivative:
\begin{equation}
\frac{\partial\rho u_\alpha u_\beta}{\partial t} = -u_\alpha\frac{\partial p}{\partial x_\gamma} \delta_{\beta\gamma}  -u_\beta\frac{\partial p}{\partial x_\gamma} \delta_{\alpha\gamma} - \frac{\partial}{\partial x_\gamma}\left( u_\alpha u_\beta u_\gamma \right),
\label{eq: drhouu/dt}
\end{equation}
and Eq.~(\ref{eq: Mass from Euler}) for the temporal derivative of the density: 
\begin{equation}
-\frac{\partial \rho c_s^2}{\partial t}\delta_{\alpha\beta} = \rho c_s^2 \frac{u_\gamma}{\partial x_\gamma} \delta_{\alpha\beta} + u_\gamma \frac{ \partial \rho c_s^2}{\partial x_\gamma} \delta_{\alpha\beta}.
\label{eq: drho/dt}
\end{equation}
In Eq.~(\ref{eq: tempo der a2eq}), the time derivative of the pressure is found from the hybrid energy part. Indeed, if one compute the partial derivative of the entropy  Eq.~(\ref{eq: Entropie expression}) with respect to the pressure and density, and replaces their expression in Eq.~(\ref{eq: Entropy eq}), the pressure equation is retrieved and reads:
\begin{equation}
\frac{\partial p}{ \partial t}\delta_{\alpha\beta} = - u_\gamma\frac{\partial p}{ \partial x_\gamma}\delta_{\alpha\beta}  - \gamma_g p\frac{\partial u_\gamma}{\partial x_\gamma}\delta_{\alpha\beta}.
\label{eq: dp/dt}
\end{equation}
This equation does not includes the viscous and diffusive terms since the dynamic viscosity is assumed to be $\mu \simeq \mathcal{O}\left(\tau\right)$ and thus $\lambda = \mu/ \left(c_p \mathrm{Pr}\right) \simeq \mathcal{O} \left(\tau\right)$ for the heat conductivity.
Inserting now Eq.~(\ref{eq: tempo der a2eq}) in the expression of $ a^{1}_{\alpha\beta}$ in Eq.~(\ref{eq: Pi inter}) gives:
\begin{equation}
\begin{split}
a^{1}_{\alpha\beta} =& - \tau \left( -E_{1,\alpha\beta} + \frac{\partial a^{eq}_{\alpha\beta\gamma}}{\partial x_\gamma} -\gamma_g p\frac{\partial u_\gamma}{\partial x_\gamma}\delta_{\alpha\beta} \right.\\ 
&+ \rho c_s^2\left( \frac{\partial u_\alpha}{\partial x_\beta} + \frac{\partial u_\beta}{\partial x_\alpha} + \frac{\partial u_\gamma}{\partial x_\gamma}\delta_{\alpha\beta} \right) \\
& + u_\alpha \frac{\partial}{\partial x_\beta}\left(\rho c_s^2-p \right)  + u_\beta \frac{\partial}{\partial x_\alpha}\left(\rho c_s^2-p \right)  \\
&\left. + u_\gamma \frac{\partial}{\partial x_\gamma}\left(\rho c_s^2-p \right) \delta_{\alpha\beta} - \frac{\partial}{\partial x_\gamma} \left( \rho u_\alpha u_\beta u_\gamma  \right) \right).
\end{split}
\label{eq:Pi neq inter}
\end{equation}
On the other side, supposing that the third-order equilibrium Hermite moment is perfectly recovered by the lattice, the following relation on $a^{eq}_{\alpha\beta\gamma}$ reads:
\begin{equation}
\begin{split}
\frac{\partial  a^{eq}_{\alpha\beta\gamma}}{\partial x_\gamma}  =& p \left( \frac{\partial u_\alpha}{\partial x_\gamma}\delta_{\beta\gamma} + \frac{\partial u_\beta}{\partial x_\gamma}\delta_{\alpha\gamma} + \frac{\partial u_\gamma}{\partial x_\gamma}\delta_{\alpha\beta} \right) \\
&-\rho c_s^2 \left( \frac{\partial u_\alpha}{\partial x_\gamma}\delta_{\beta\gamma} + \frac{\partial u_\beta}{\partial x_\gamma}\delta_{\alpha\gamma} + \frac{\partial u_\gamma}{\partial x_\gamma}\delta_{\alpha\beta} \right) \\
&+u_\alpha\frac{\partial}{\partial x_\gamma} \left(p-\rho c_s^2\right)\delta_{\beta\gamma}+u_\beta\frac{\partial}{\partial x_\gamma} \left(p-\rho c_s^2\right)\delta_{\alpha\gamma} \\
&+u_\gamma\frac{\partial}{\partial x_\gamma} \left(p-\rho c_s^2\right)\delta_{\alpha\beta}+\frac{\partial}{\partial x_\gamma} \left( \rho u_\alpha u_\beta u_\gamma \right),
\end{split}
\end{equation}
and once inserted in Eq.~(\ref{eq:Pi neq inter}) eventually yields:
\begin{equation}
\begin{split}
a^{1}_{\alpha\beta}  =& - \tau \left( -E_{1,\alpha\beta} + p \left( \frac{\partial u_\alpha}{\partial x_\beta} + \frac{\partial u_\beta}{\partial x_\alpha} - \left(\gamma_g-1\right) \frac{\partial u_\gamma}{\partial x_\gamma}\delta_{\alpha\beta} \right) \right).
\end{split}
\label{eq:Pi neq inter2}
\end{equation}
From this point, the correction terms $E_{1,\alpha\beta}$ is needed to obtain the correct viscous stress tensor associated to the Navier-Stokes equations.
Indeed, in order to obtain Eq.~(\ref{eq:Pi neq inter2}) , the third-order Hermite equilibrium moment (Eq.~(\ref{eq: a3})) has been supposed to be exactly recovered by the lattice which is not the case for the $D2Q9$. Thus in the light of Eq.~(\ref{eq: closure a3 Q9 Q27}), this ultimately settles the form of the correction term at a macroscopic level as:
\begin{equation}
\begin{split}
E_{1,\alpha\beta} =&  - \frac{\partial}{\partial x_\alpha}  \left( a^{eq}_{\alpha\alpha\alpha} \right)  \delta_{\alpha\beta}.
\end{split}
\end{equation}

At last, the stress tensor, now free of the lattice closure error reads:
\begin{equation}
\begin{split}
a^{1}_{\alpha\beta}  =& - \tau p \left( \frac{\partial u_\alpha}{\partial x_\beta} + \frac{\partial u_\beta}{\partial x_\alpha} - \left(\gamma_g-1\right) \frac{\partial u_\gamma}{\partial x_\gamma}\delta_{\alpha\beta} \right),
\end{split}
\label{eq:Pi neq final}
\end{equation}
in which, by identification to the Navier-Stokes equations, the dynamic viscosity is found to be $\mu=\tau p$.
The macroscopic set of equations recovered by the HLBM model is then:
\begin{subequations}
\begin{align}
\label{eq: Mass from NS}&\frac{\partial \rho}{\partial t} + \frac{\partial \rho u_\alpha}{\partial x_\alpha} = 0 \\
\label{eq: QDM from NS}&\frac{\partial \rho u_\alpha}{\partial t} + \frac{\partial }{\partial x_\beta} \left(\rho u_\alpha u_\beta + p \delta_{\alpha\beta} \right)  = -\frac{\partial a^{1}_{\alpha\beta} }{\partial x_\beta} \\
\label{eq: entrop from NS}&\frac{\partial s}{ \partial t} + u_\alpha \frac{\partial s}{ \partial x_\alpha}   = - \frac{1}{ \rho \theta}\frac{\partial }{ \partial x_\alpha} \left(- \lambda  \frac{\partial \theta}{ \partial x_\alpha}  \right) + \frac{\Phi}{ \rho \theta},
\end{align}
\label{eq: NS system}
\end{subequations}
with
\begin{equation}
p=\rho c_s^2\theta \qquad \mathrm{and} \qquad s=c_v\ln\left( \frac{\theta}{\rho^{ \left(\gamma_g-1\right) }} \right),  
\end{equation}
and the Mayer's relation $c_p - c_v = c_s^2$ giving the expression of the dimensionless heat capacities:
\begin{equation}
c_v = \frac{c_s^2}{\left(\gamma_g-1\right)} \qquad \mathrm{and} \qquad c_p = \frac{\gamma_g c_s^2}{\left(\gamma_g-1\right)}.
\end{equation}

\subsection{\label{subsec: Speed of sound}Speed of sound}
In the standard isothermal case, the athermal speed of sound is fixed and imposed by the lattice constant $c=c_s$ (which reads $c^*=\sqrt{rT_r}$ for its dimensional counterpart). This fixes the choice of $rT_r$ to recover the desired acoustic speed. In the present case, and in the light of the previous section (Sec.~\ref{sec:Macro equation}), the Hybrid DVBE is able to solve the Navier-Stokes-Fourier equations with the perfect gas equation of state (e.o.s) $p=\rho c_s^2 \theta$. Inserting this e.o.s in the entropy expression (Eq.~(\ref{eq: Entropie expression})), one can write the pressure as:
\begin{equation}
p = \rho^{\gamma_g} c_s^2 \exp \left( \frac{s\left( \gamma_g - 1 \right)}{c_s^2}\right).
\end{equation}
The dimensionless isentropic speed of sound can be now expressed as:
\begin{equation}
c^2=\frac{\partial p}{\partial \rho}|_s =  \gamma_g \frac{p}{\rho} = \gamma_g c_s^2 \theta,
\label{eq: speed of sound}
\end{equation}
which once redimensionalized by the arbitrary length $\Delta x$ and time $\Delta t$ reads:
\begin{equation}
c^*=\sqrt{\gamma_g rT}. 
\label{eq: speed of sound 2}
\end{equation}
A temperature and heat capacity ratio dependent acoustic speed is thus well recovered by this Hybrid DVBE model.

\section{\label{sec: Improvements} Extended HLBM for high Mach number compressible flows}

In this section, the improvements made to the original model~\cite{Feng2019}, which allow it to be extended to supersonic regimes, are presented. First, the generic spatiotemporal discretization of the hybrid DVBE is presented, leading to the well known collide and stream algorithm. Secondly, an enhanced correction term is derived, engendering a traceless viscous stress tensor. The discretization scheme for the correction term is described, as well as the standard lattice closure term, where a new tailored scheme is proposed. Third, the hybrid recursive regularized collision operator~\cite{Jacob2019} is presented, and slightly adapted according to the modification brought by the new correction term. Fourth, the discretization of the hybrid energy part is detailed, and a new numerical scheme for the heat production term, combining finite difference and LBM, is proposed. Finally, the algorithm of the present HLBM is detailed with a brief discussion on the Courant–Friedrichs–Lewy number.

\subsection{\label{subsec: Time space discr DVBE} Space/time discretization of the DVBE}

The corrected discrete velocity Boltzmann BGK equation:
\begin{equation}
\frac{\partial f_i}{\partial t} + c_{i,\alpha}\frac{\partial f_i}{\partial x_\alpha} = -\frac{1}{\tau}\left( f_i - f_i^{eq} \right) + \psi_i,
\end{equation}
can be solved numerically by discretizing the space and time of the discrete probability functions $f_i$ on a Cartesian mesh, in a particular way. This consists in integrating between $t$ and $t+\Delta t$ the LHS linear convection term along the characteristic $c_{i,\alpha}$, whereas the trapezoidal rule ensures a second-order accuracy in time for the RHS collision and force terms integration. This strategy leads to:
\begin{equation}
\begin{split}
f^+_i - f_i = &-\frac{\Delta t}{2\tau}\left( f_i^+ - f_i^{eq,+} + f_i - f_i^{eq} \right)\\
& +\frac{\Delta t}{2}\left( \psi_i^+ - \psi_i \right) + \mathcal{O}\left(\Delta t^2, \Delta x^2 \right),
\label{eq: LBM step 1}
\end{split}
\end{equation}
where the superscript $+$ denotes the solution at $\left(x+c_{i,\alpha}\Delta t, t+\Delta t \right)$ with $\Delta t$ and $\Delta x = c_{i,\alpha}\Delta t$ referring respectively to the time and space step. To remove the implicit character of the formulation, a change of variable compliant with the conservation of mass and momentum is adopted~\cite{he1998discrete,he1998novel,dellar2013interpretation} and reads:
\begin{equation}
\bar{f_i} = f_i + \frac{\Delta t}{2\tau} \left(f_i - f^{eq}_i\right) - \frac{\Delta t}{2} \psi_i,
\label{eq:change of variable 1}
\end{equation}
which once plugged back in Eq.~(\ref{eq: LBM step 1}) gives:
\begin{equation}
\begin{split}
\bar{f_i}^+ = & f_i  -\frac{\Delta t}{2\tau}\left( f_i - f_i^{eq} \right) + \frac{\Delta t}{2} \psi_i.
\label{eq: LBM step 2}
\end{split}
\end{equation}
Inverting now Eq.~(\ref{eq:change of variable 1}), one obtains:
\begin{equation}
\begin{split}
f_i = & \frac{2\tau}{2\tau+\Delta t} \left( \bar{f_i} + \frac{\Delta t}{2\tau}  f_i^{eq} + \frac{\Delta t}{2} \psi_i\right),
\label{eq: LBM trick 1}
\end{split}
\end{equation}
and once inserted in Eq.~(\ref{eq: LBM step 2}), after some algebra, finally gives:
\begin{equation}
\begin{split}
\bar{f_i}^+ = & \bar{f_i}  -\frac{\Delta t}{\bar{\tau}}\left( \bar{f_i} - f_i^{eq} \right) + \frac{2\bar{\tau} - \Delta t}{2\bar{\tau}}\Delta t \psi_i.
\label{eq: LBM BGK}
\end{split}
\end{equation}
This is the collide-and-stream corrected LBM-BGK equation~\cite{Feng2018} with $\bar{\tau}=\tau+\Delta t/2$ the new relaxation time induced by the change of variable. The function $\bar{f_i}$ is then computed in time and space following a collision step and a streaming step (i.e. a direct propagation from node to node along $c_{i,\alpha}$). Computing the moments of the distribution function yields the mass at the zeroth order:
\begin{equation}
\rho \equiv \sum_{i=0}^{m-1}\bar{f_i} = \sum_{i=0}^{m-1}f_i^{eq} ,
\label{eq: rhoalgo}
\end{equation} 
and the momentum at first-order:
\begin{equation}
\rho u_\alpha \equiv \sum_{i=0}^{m-1} c_{i,\alpha} \bar{f_i} = \sum_{i=0}^{m-1} c_{i,\alpha} f_i^{eq},
\label{eq: ualgo}
\end{equation} 
which are still left invariant by collision by construction thanks to the feature of Eq.~(\ref{eq:change of variable 1}).

\subsection{\label{subsec: Corr terms enhancement} Correction term enhancement and discretization}

Beginning with the viscous stress tensor found after the Chapmann-Enskog development in Sec.~\ref{sec:Macro equation}:
\begin{equation}
\begin{split}
a^{1}_{\alpha\beta}  =& - \tau p \left( \frac{\partial u_\alpha}{\partial x_\beta} + \frac{\partial u_\beta}{\partial x_\alpha} - \left(\gamma_g-1\right) \frac{\partial u_\gamma}{\partial x_\gamma}\delta_{\alpha\beta} \right).
\end{split}
\label{eq:Pi neq inter2.1}
\end{equation}
Although this present term is free of the $D2Q9$ lattice closure error, it is still not correct if the heat capacity ratio $\gamma_g$ differs from the monatomic value. This relates to a second error, the latter being inherent to the hybrid method. This error is a direct consequence of the monatomic modelling of LBM where, $\gamma_{lb}=\left(D+2\right)/D$ with $D$ referring simply to the dimension, while the heat capacity ratio $\gamma_g$ in the hybrid energy part can be chosen freely. This defect modifies the trace of the viscous stress tensor and introduces additional bulk viscosity impacting the dissipation rate of the acoustic waves. Following the same procedure as in Sec.~\ref{sec:Macro equation} to find $E_{1,\alpha\beta}$, the second correction term $E_{2,\alpha\beta}$ is settled at the macroscopic level as:
\begin{equation}
E_{2,\alpha\beta} =p\left(\frac{D+2}{D}-\gamma_g\right) \frac{\partial u_\gamma}{\partial x_\gamma}\delta_{\alpha\beta}.
\label{eq: E2 corr term}
\end{equation}
Thus, the enhanced correction term at the Boltzmann level is given by:
\begin{equation}
\psi_i = w_i\frac{\mathcal{H}_{i,\alpha\beta}}{2c_s^4} \left( E_{1,\alpha\beta} + E_{2,\alpha\beta} \right),
\label{eq: Psialgo}
\end{equation}
leading to the error-free viscous stress tensor:
\begin{equation}
\begin{split}
a^{1}_{\alpha\beta}  =& - \tau p \left( \frac{\partial u_\alpha}{\partial x_\beta} + \frac{\partial u_\beta}{\partial x_\alpha} - \frac{2}{D} \frac{\partial u_\gamma}{\partial x_\gamma}\delta_{\alpha\beta} \right). 
\end{split}
\label{eq:Pi neq final2}
\end{equation}

In the present work, the standard second-order centered finite difference is used for the discretization of $E_{2,\alpha\beta}$.
However, for the $E_{1,\alpha\beta}$ correction term, an upwind biased finite difference scheme is adopted and reads for the $x$ direction:
\begin{equation}
E_{1,xx} = \frac{\left[ 1 + \mathrm{sgn}(u_x) \right]}{2} \frac{ \Gamma_{i} - \Gamma_{i-1}}{\Delta x} + \frac{\left[ 1 - \mathrm{sgn}(u_x) \right]}{2} \frac{ \Gamma_{i+1} - \Gamma_{i}}{\Delta x}.
\end{equation}
Here, $\Gamma=\rho u_x\left(1-\theta - u_x^2 \right)$ and $i$ represents the spatial position on the grid. $\mathrm{sgn}(u_x)$ simply denotes the sign of $u_x$. This discretization differs from the centered scheme used by the original model~\cite{Feng2019}, and has been found non to be much more stable and mandatory for high Mach number cases as demonstrated in the next section~\ref{sec: Validation of the improvements}.

\subsection{\label{subsec:HRR}Hybrid Recursive Regularised collision operator}

A Hybrid Recursive Regularised (HRR)~\cite{Jacob2019} collision operator has been chosen. It offers fewer degrees of freedom compared to the Multiple Relaxation Time (MRT) operators and its efficiency and robustness in the HLBM case has already been demonstrated~\cite{Feng2019}. This HRR collision operator is based on a partial reconstruction of the off-equilibrium part of the pre-collision distribution function $\bar{f}_i^1$ using finite differences. This permits a significant gain in terms of numerical stability, partially filtering out non-hydrodynamics modes, and suppressing their interactions with the other physical modes~\cite{Astoul2019}.
Just before the collision step, the off-equilibrium part of the distribution functions is regularized according to:
\begin{equation}
\bar{f}_i^{1} = w_i \sum_{n=2}^{N} \frac{1}{n!\left(c_s^2\right)^n}\boldsymbol{\bar{a}}^{1,\left(n\right)} : \boldsymbol{\mathcal{H}}_i^{\left(n\right)},
\end{equation}
and reconstructed following $\bar{f_i} = f_i^{eq} + \bar{f_i^{1}}$, with $\bar{f_i^{1}}=\bar{f_i} - f_i^{eq} + \Delta t \psi_i /2 $~\cite{Feng2019}. This procedure is equivalent to re-writing Eq.~(\ref{eq: LBM BGK}) as:
\begin{equation}
\bar{f_i}^+ = f_i^{eq} + \left(1 - \frac{\Delta t}{ \bar{\tau}} \right) \bar{f_i^{1}} + \frac{\Delta t }{2}\psi_i.
\label{eq: streamalgo}
\end{equation} 
Here it is worth noting that mass and momentum are collision invariants, so that the off-equilibrium reconstruction starts from the second-order moment, i.e. the viscous stress tensor.
In the HRR framework, this tensor is reconstructed as:
\begin{equation}
\bar{a}^1_{\alpha,\beta} = \sigma \bar{a}^{1,\mathrm{PR}}_{\alpha,\beta} + \left(1-\sigma\right) \bar{a}^{1,\mathrm{FD}}_{\alpha,\beta},
\label{eq: a1sigalgo}
\end{equation}
where the bar notation stands for the moment of the LBM system after the change of variable. Above, $\bar{a}^{1,\mathrm{PR}}_{\alpha,\beta}$ is the standard second-order off-equilibrium moment obtained by projection~\cite{Latt2006}:
\begin{equation}
\bar{a}^{1,\mathrm{PR}}_{\alpha,\beta} = \sum_{i=0}^{m-1} \mathcal{H}_{i,\alpha\beta} \left(\bar{f_i} - f_i^{eq} + \frac{\Delta t}{2} \psi_i \right),
\end{equation}
and $a^{1,\mathrm{FD}}_{\alpha,\beta}$ is reconstructed using second-order centered finite differences, discretizing the viscous stress tensor Eq.~(\ref{eq:Pi neq final2}):
\begin{equation}
\bar{a}^{1,\mathrm{FD}}_{\alpha,\beta} = - \frac{\bar{\tau}}{\tau}\mu \left(\frac{\partial u_\alpha}{\partial x_\beta} + \frac{\partial u_\beta}{\partial x_\alpha} - \frac{2}{D} \frac{\partial u_\gamma}{\partial x_\gamma} \delta_{\alpha\beta} \right),
\label{eq: a1FD}
\end{equation}
and has been retrieved error free thanks to the new $E_{2,\alpha\beta}$ correction term.
The coefficient $\bar{\tau}/\tau$ appears naturally due to the change of variable impacting the non-conserved moments~\cite{Dellar2001}.
Once the second-order off-equilibrium moment is determined using the previous procedure, one can retrieved the higher off-equilibrium moment using the recurrence formula of~\cite{Coreixas2017}.
This is done up to the fourth order, and only for the moments supported by the lattice which read for the $D2Q9$:
\begin{eqnarray}
 &  & \bar{a}^{1}_{xxy} \simeq  u_y \bar{a}^1_{xx} + 2u_x \bar{a}^{1}_{xy},  \\
 &  & \bar{a}^{1}_{yyx} \simeq   u_x \bar{a}^1_{yy} + 2u_y \bar{a}^{1}_{xy},  \\
 &  & \bar{a}^{1}_{xxyy}\simeq   2\left(u_{x} \bar{a}_{y y x}^{1}+u_{y} \bar{a}_{x x y}^{1}\right) + \left[c_{s}^{2}(\theta-1)-u_{x}^{2}\right] \bar{a}_{ y y}^{1} \nonumber \\
 &  & \qquad\quad\; + \left[c_{s}^{2}(\theta-1)-u_{y}^{2}\right] \bar{a}_{x x}^{1}-4 u_{x} u_{y} \bar{a}_{x y}^{1} \\
 \nonumber
\end{eqnarray}

\subsection{\label{subsec: Entrop discr} Entropy equation discretization}

Here, the discretization in time and space of the hybrid energy part is presented. The entropy equation:
\begin{equation}
\frac{\partial s}{ \partial t} + u_\alpha \frac{\partial s}{ \partial x_\alpha}   = - \frac{1}{ \rho \theta}\frac{\partial }{ \partial x_\alpha} \left(- \lambda  \frac{\partial \theta}{ \partial x_\alpha}  \right) +   \frac{\Phi}{ \rho \theta}, 
\label{eq: sequaalgo}
\end{equation}
is composed of an advection part on the LHS, and a diffusion and production terms on the RHS. These RHS terms correspond respectively to the Fourier diffusion and to the viscous heat production.
Similarly to~\cite{Feng2019}, the convective flux is computed using a MUSCL scheme with a Van-Albada flux limiter. A standard second-order centered finite difference scheme is employed for the Fourier term. However in the present model, the viscous heat production term $\Phi$, is partially computed using finite differences and using $\bar{a}^1_{\alpha\beta}$ computed by the LBM part. This term reads:
\begin{equation}
\begin{split}
\Phi &= \mu\left( \frac{\partial u_\alpha}{\partial x_\beta} + \frac{\partial u_\beta}{\partial x_\alpha} - \frac{2}{D} \frac{\partial u_\gamma}{\partial x_\gamma} \delta_{\alpha\beta} \right) \frac{\partial u_\alpha}{\partial x_\beta} \\
&=-\frac{\tau}{\bar{\tau}}\bar{a}^1_{\alpha\beta} \frac{\partial u_\alpha}{\partial x_\beta}.
\end{split}
\label{eq:viscous heat discret}
\end{equation}
In addition to improving the consistency between the LBM and the energy system, this strategy allows for the reduction of the number of operations to compute this tedious term. Indeed, in the regularised framework, only the $\bar{a}^1_{\alpha\beta}$ tensor needs to be stored in memory since the distribution function $\bar{f}_i$ is entirely reconstructed before the collision step. Thus, since this tensor is available, it can be used without any additional cost to model $\Phi$ by a multiplication with the velocity gradient, in Eq.~(\ref{eq:viscous heat discret}), modeled with a standard second order finite difference scheme. Moreover, if the $\bar{a}^1_{\alpha\beta}$ tensor was not traceless thanks to the $E_{2,\alpha\beta}$ term in Eq.~(\ref{eq: E2 corr term}), it is worth noting that a spurious heat production would bias any thermal solutions.
Finally, this discretization has been successfully tested on smooth flow (see Sec.~\ref{sec:couette}), however due to the centered nature of the velocity gradient, in case of discontinuities it is preferable to model it using only finite difference scheme. 
Finally, the standard explicit Euler scheme is used for the time marching procedure.

\subsection{\label{subsec: CFL and HLBM algo} CFL number and HLBM algorithm}

In most of the cases, the operating point of an explicit numerical scheme is characterized by the Courant-Friedrichs-Lewy (CFL) number. Once the CFL restriction is known, this permits the choice of the appropriate time step value in function of the local mesh size and the maximum expected fluid velocity. The CFL number in the compressible Navier-Stokes framework is expressed as the ratio of the maximal physical information speed, i.e. $c^*\left(\mathrm{Ma}+1\right)$, over the numerical speed $\Delta x / \Delta t$.
In the standard case, the athermal LBM speed of sound is imposed by the acoustic scaling $c_s \Delta x/\Delta t = \sqrt{rT_r}$, fixing the choice of the time step to recover the desired acoustic speed at $c^*=\sqrt{rT_r}$ (which reads $c=c_s$ for its dimensionless counterpart). However in the present case, this HLBM model is now function of the fluid temperature and independent of the reference temperature $T_r$. Then the time step can be freely chosen depending on the value of $T_r$, without any impact on the desired acoustic speed. 
Replacing the time step by its expression, this results in:
\begin{equation}
\mathrm{CFL} = \left( \mathrm{Ma} + 1 \right) \sqrt{\gamma_g c_s^2 \theta}.
\label{eq:CFL}
\end{equation}
As the Mach number and the heat capacity ratio belong to the physics of the problem, the CFL number is adjusted only using the numerical parameter $T_r$ enabling to reduce or increase the time step, to tailor stability and/or convergence. However, one has to keep in mind that the present HLBM scheme is explicit, and must obey the upper bound restriction of $\mathrm{CFL}<1$. Thus the speed convergence through $T_r$ is constrained by the relation: 
\begin{equation}
\theta_{max} <\frac{1}{\gamma \left(\mathrm{Ma} + 1\right)^2 c_s^2}. 
\label{eq: CFL constraint}
\end{equation}

Finally, the present HLBM algorithm reads chronologically:
\begin{algorithm}[H]
\begin{spacing}{1.4}

\textbf{Input:}

1 : $\left[\,\rho,\, u_\alpha,\, \theta\,\right]^{(0)}$
\\
\\
\textbf{Initialization:}

2 : Compute $f_i^{eq,(0)}$ from $\left[\,\rho,\, u_\alpha,\, \theta\,\right]^{(0)}$ ; Eq.~(\ref{eq:Equilibrium})

3 : Compute $\bar{a}_{\alpha\beta}^{1,(0)}=\bar{a}_{\alpha\beta}^{1,FD}$ from $\left[\,\rho,\, u_\alpha,\, \theta\,\right]^{(0)}$ ; Eq.~(\ref{eq: a1FD})

4 : Compute  $\psi^{(0)}_i$ from $\left[\,\rho,\, u_\alpha,\, \theta\,\right]^{(0)}$ ; Eq.~(\ref{eq: Psialgo})
\\
\\
\textbf{for} $n=1\rightarrow N$ \textbf{do:}

5 : Compute $\bar{f}_i^{+}$ from $\left[\,f_i^{eq},\,\bar{a}_{\alpha\beta}^{1},\,\psi_i \,\right]^{(n)}$ ; Eq.~(\ref{eq: streamalgo})

6 : Compute $\left[\,\rho,\, u_\alpha\,\right]^{(n+1)}$ from $\bar{f}_i^{+}$ ; Eq.~(\ref{eq: rhoalgo}) and Eq.~(\ref{eq: ualgo})

7 : Compute $s^{(n+1)} $ from $\left[\, \rho,\, u_\alpha,\, \theta,\, \bar{a}^{1}_{\alpha\beta}\, \right]^{(n)}$ ; Eq.~(\ref{eq: sequaalgo}) 

8 : Compute $\theta^{(n+1)}$ from $\left[\,s,\,\rho\,\right]^{(n+1)}$ ; Eq.~(\ref{eq: Entropie expression})

9 : Compute  $\psi^{(n+1)}_i$ from $\left[\,\rho,\,u_\alpha,\,\theta\,\right]^{(n+1)}$  ; Eq.~(\ref{eq: Psialgo})

10 : Compute $f_i^{eq,(n+1)}$ from $\left[\,\rho,\, u_\alpha,\, \theta\, \right]^{(n+1)}$ ; Eq.~(\ref{eq:Equilibrium})

11 : Compute $\bar{a}_{\alpha\beta}^{1,(n+1)}$ from $\bar{f}_i^{+}$, $\left[\,f_i^{eq},\, \psi_i,\, \rho,\, u_\alpha,\, \theta\, \right]^{(n+1)}$  ; Eq.~(\ref{eq: a1sigalgo}) 

\end{spacing}
\caption{HLBM}
\label{LBM Algo}
\end{algorithm}
\noindent where $n$ and $N$ stand respectively for the $n^{th}$ and final time step. It is worth noting that, the present scheme permits the natural initialization of the off-equilibrium part of $f_i$, which might be of importance for initialisation-sensitive flows~\cite{Kruger_Book_2017} adversely affected by an equilibrium state initialization. Moreover, this algorithm slightly differs from the one of~\cite{Feng2019}, in a sense that it does not take into account the two-step temporal integration for the energy equation. Tests have been done with and without this integration, and no substantial improvements in terms of stability or accuracy have been noticed.

\section{\label{sec: Validation of the improvements} Numerical validations of the improved HLBM}

In this section, the proposed improvements brought to the HLBM scheme are assessed in different configurations and compared to the original model of Feng $et$ $al.$~\cite{Feng2019}. The improvements consist of: (1) a new discretization of the lattice closure correction term, (2) a corrected viscous stress tensor that takes into account polyatomic gases, and (3) a novel discretization of the viscous heat production term fitting with the regularized formalism. The aim is to verify that, thanks to the improvements, all the physical phenomena related to the Navier-Stokes Fourier equations are correctly retrieved, both in the subsonic regime (and without regression compared to the original model), as well as in the supersonic regime. 
The simulations show the restriction of the original model to subsonic cases, while the new enhancements of the present model make it possible to overcome this limitation and extend the range of application of the HLBM to supersonic cases.

\subsection{\label{sec:shear wave}Viscosity computation}

Here the kinematic viscosity $\nu$ will be computed from the response of the numerical system to a shear wave initialization. The domain is periodic and the initialization reads:
\begin{equation}
p = p_\infty, \: u_x=a_0\sin\left(2\pi y /L_y \right), \: u_y=u_\infty, \: \: T=T_\infty, \nonumber
\label{eq: Shear_init}
\end{equation}
where $T_\infty = 300\,\mathrm{K}$, $p_\infty=101325\,\mathrm{Pa}$ and the amplitude of the perturbation has been arbitrary taken to $a_0=20\, \mathrm{m.s}^{-1}$. The domain length is $L_y=1\, \mathrm{m}$ and the velocity propagation of the perturbation is $u_\infty=\mathrm{Ma}\, c_\infty$, with $c_\infty=\sqrt{\gamma_g r T_\infty}$ the speed of sound and $\gamma_g=1.4$ in the present case.
Similarly to~\cite{saadat2019} the setup is quasi one dimensional, $200$ points per wavelength were considered, and the wave has one period along the $y$ direction. The temporal decay of this wave is directly linked to the kinematic viscosity and the analytical solution for the velocity reads~\cite{Chu1957,shan2007general}:
\begin{equation}
max\left(u_x\right)=a_0\exp\left(-\nu \left(\frac{2\pi}{L_y}\right)^2 t \right).
\label{eq: Shearwave}
\end{equation}
Thus tracking the maximum of $u_x$ along time and fitting the curve using a least square method yields the experimental value of the viscosity $\nu$.

\begin{figure}[!t]
\centering
\includegraphics[width=1.0\columnwidth]{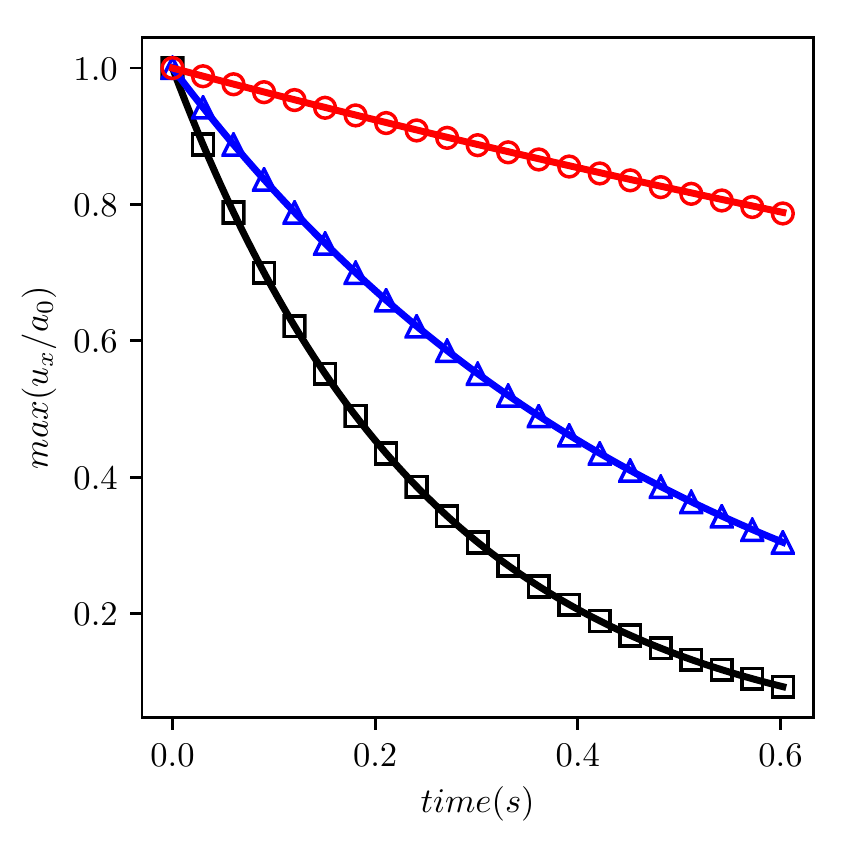}
\caption{Velocity decay of a shear wave along the time for different viscosities at $\mathrm{Ma}=1.5$. The analytical solution is plotted in solid line and the symbols correspond to the present HLBM model for $\nu=0.01 \,\mathrm{m^2.s}^{-1}$ \protect\RedCircle; $\nu=0.05\,\mathrm{m^2.s}^{-1}$   \protect\BlueTriangle; and $\nu=0.1\,\mathrm{m^2.s}^{-1}$  \protect\BlackSquare. The numerical parameters were $\sigma=0$ and $T_{r}=800\,\mathrm{K}$.}
\label{fig: Shearwave}
\end{figure}

Fig.~\ref{fig: Shearwave} shows a good match of the present model (symbols) with the theory Eq.~(\ref{eq: Shearwave}) (solid lines), for different viscosities at a supersonic advection regime of $\mathrm{Ma}=1.5$.
More results in terms of Mach number and viscosity can be found in Table~\ref{table: Shearwave}, where the present model is compared to the original model~\cite{Feng2019} in terms of relative error. 
From this table, it worth noting that the present HLBM model ensures the retrieval of the correct kinematic viscosity from low to high Mach numbers. For the low Mach number case, similar results have been found compared to the original model. However, this last turns out to be unstable for Mach numbers greater or equal to unity restricting its application range to subsonic cases only. The results clearly show the enhancement of this scheme in terms of robustness without regressions in terms of accuracy, extending the validity of the model for this case, to supersonic regime.

\begin{table}
%\footnotesize
\small
\begin{tabular}{|c|c|c||c|c|}
\multicolumn{1}{c}{} & \multicolumn{2}{c}{$\nu=0.1\,\mathrm{m^2.s}^{-1}$} & \multicolumn{2}{c}{$\nu=0.05\,\mathrm{m^2.s}^{-1}$}\tabularnewline
\hline 
\hline 
Ma & $\epsilon \left(\mathrm{P}\right)$ \% & $\epsilon \left(\mathrm{O} \right)$ \% & $\epsilon \left(\mathrm{P} \right)$ \% & $\epsilon \left(\mathrm{O}\right)$ \%\tabularnewline
\hline 
$0.5$ & $4.72.10^{-4}$ & $4.72.10^{-4}$  & $1.24.10^{-3}$ & $1.24.10^{-3}$ \tabularnewline
\hline 
$1.0$ & $6.84.10^{-4}$ & $\varnothing$ & $1.67.10^{-3}$  &$\varnothing$ \tabularnewline
\hline 
$1.5$ & $7.12.10^{-4}$ & $\varnothing$  & $1.63.10^{-3}$ &$\varnothing$ \tabularnewline
\hline 
\end{tabular}\caption{Relative error on the viscosity associated to the original model~\cite{Feng2019} noted O, and the present model noted P. The symbol $\varnothing$ means that the computation diverged. For the shear wave the numerical parameters were $\sigma=0$ and $T_{r}=800 \,\mathrm{K}$.}
\label{table: Shearwave}
\end{table}

\subsection{\label{sec:covo}Supersonic isentropic vortex}

The convected isentropic vortex is an Euler benchmark that can be used to assess numerical schemes in term of dispersion, dissipation and robustness. Although the initialization of this vortex is based on the Euler equations, it has been intensively used to bench different LBM schemes solving the Navier-Stokes equation~\cite{saadat2019,Frapolli2016a}. Despite the lack of thermal and viscous diffusive effects in the solution, this case can still provide useful informations on the dispersion and stability properties thanks to its analytical solution. This exact solution is simply the advection of the initial condition over the time. The initialization can be seen as a small perturbation of the mean flow $\rho_{\infty}, T_\infty, u_\infty$ which reads: 

\begin{equation}
u_r(r)=0 ~ \mathrm{and}~ u_\theta(r)=c_\infty Ma_vr\exp{[(1-r^2)/2]},
\label{eq: V covo}
\end{equation}
where $u_r$ and $u_{\theta}$ are respectively the radial and tangential velocity in polar coordinates. $\mathrm{Ma}_v$ is the vortex Mach number and $r=\sqrt{ \left(x-x_c\right)^2  + \left(y-y_c\right)^2 } / R$ with $R$ the vortex radius. This vortex is considered isentropic, i.e. $p/\rho^{\gamma_g}=cst$, leading naturally to the following density field:
\begin{equation}
\rho(r)=\rho_\infty\left[1-\frac{\gamma_g-1}{2}Ma_v^2r\exp{(1-r^2)}\right]^{\frac{1}{\gamma_g-1}},
\label{eq: rho covo}
\end{equation}
and pressure fields:
\begin{equation}
p(r)=p_\infty \left( \frac{\rho\left(r\right)}{\rho_\infty} \right)^{\gamma_g}.
\label{eq: p covo}
\end{equation}
In this case, the value of the vortex velocity was chosen as $\mathrm{Ma}_v = \beta*Ma_\infty$, with $\beta=0.1$ representing the vortex intensity. $\mathrm{Ma}_\infty$ denotes the freestream Mach number. A $\left[200\times200 \right]$ mesh was considered in a one-meter periodic box, and twenty points were taken in the vortex radius. The dynamic viscosity is taken as the same order as the air value $\mu=1.10^{-5}\,\mathrm{kg.m^{-1}.s}^{-1}$, and the reference pressure and temperature are taken to $p_\infty = 101325\,\mathrm{Pa}$ and $T_\infty = 300\,\mathrm{K}$ respectively. 
The advection velocity range is taken from $\mathrm{Ma}_\infty=0.3$ to $\mathrm{Ma}_\infty=1.3$ which corresponds to typical Mach number regimes encountered in aeronautical flows i.e. approach ($Ma=0.3$), cruise ($Ma=0.8$) and local max velocity nozzle ($Ma=1.3$).
A unique $T_{r}=1478.75\,\mathrm{K}$ (defining the time step) was chosen over all Mach numbers. This $T_r$ value was calibrated on the stability of the higher Mach case, namely $\mathrm{Ma}_\infty=1.3$, and corresponds in the present case to $\mathrm{CFL}\simeq0.7$.

\begin{figure}[!t]
\centering
\includegraphics[width=1.0\columnwidth]{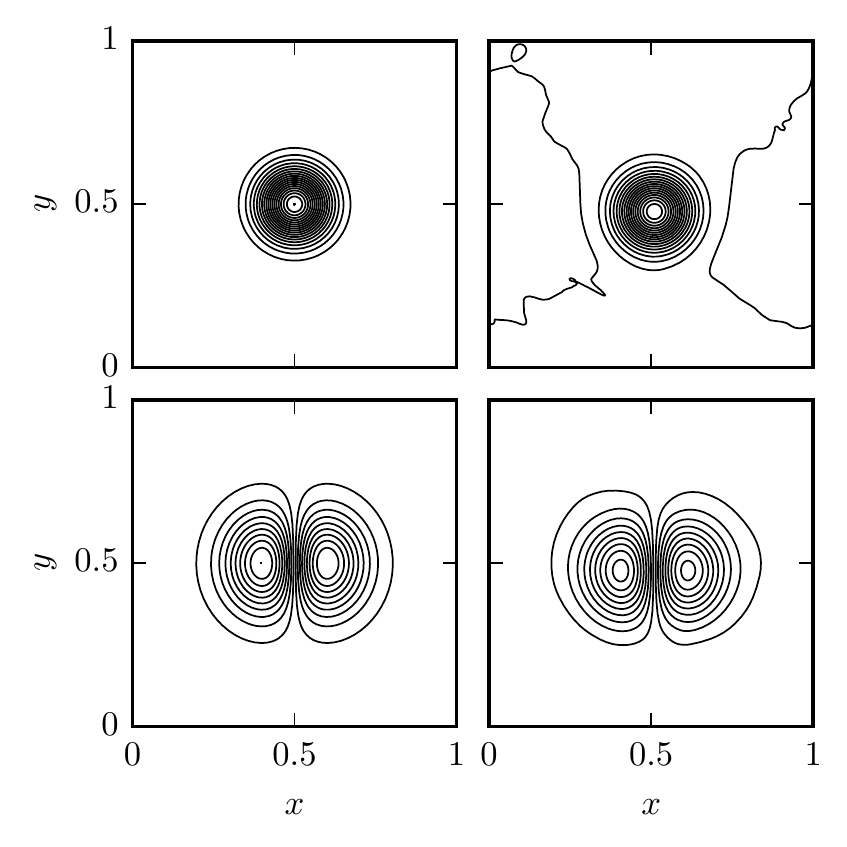}
\caption{Convected isentropic vortex at $\mathrm{Ma}_\infty=1.3$ before (left) and after (right) $50$ convective times. Iso-contour of density (top) and $y$ component of the velocity (bottom) show excellent results in terms of isotropy.}
\label{fig: COVO_isoRHO_VY}
\end{figure}
\begin{figure}[!t]
\centering
\includegraphics[width=1.0\columnwidth]{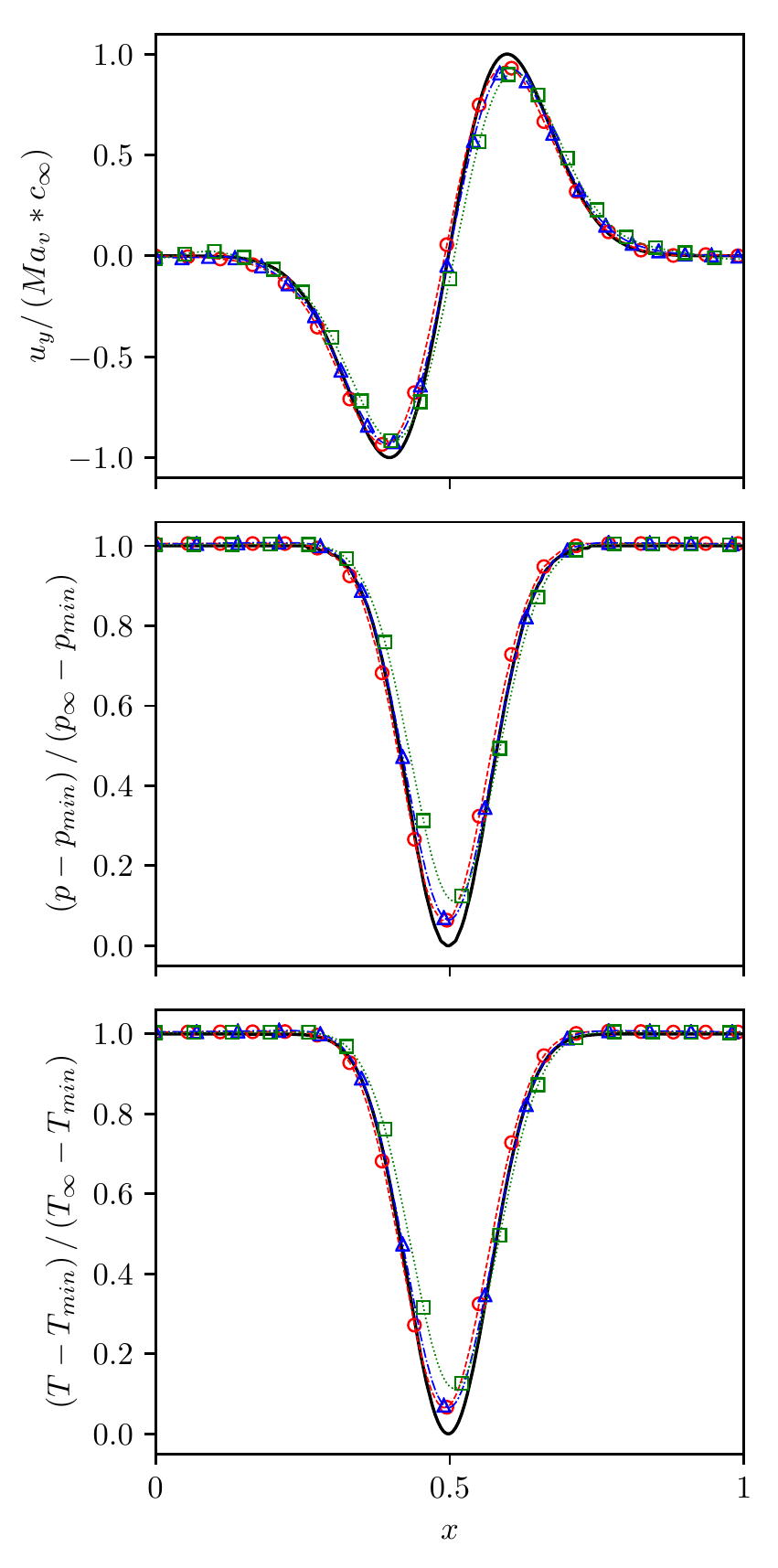}
\caption{Convected isentropic vortex for different Mach numbers after $50$ convective times at $\sigma=0$. The $min$ subscript corresponds to the minimal value at the initialization. \protect\BlackLine corresponds to analytic; \protect\RedDottedCircle corresponds to $\mathrm{Ma}_\infty=0.3$; \protect\BlueDottedTriangle to $\mathrm{Ma}_\infty=0.8$ and \protect\GreenDottedSquare to $\mathrm{Ma}_\infty=1.3$. }
\label{fig: COVO_Cut_p_VY}
\end{figure}

The velocity and density contours in Fig.~\ref{fig: COVO_isoRHO_VY}, show the respect of the isotropy of the fields by the present scheme. These results have been obtained for the highest Mach number case $\mathrm{Ma}_\infty=1.3$ with $\sigma=0$. More quantitatively, from Fig.~\ref{fig: COVO_Cut_p_VY}, one can see on different profiles, the overall agreement of the solution (symbols) with the analytic (solid lines), for three different Mach numbers, ranging from subsonic to supersonic regime. On these graphs, one can remark that the vortex is more dissipated for $\mathrm{Ma}_\infty=1.3$ than the two lower Mach number cases. This is due to the similar time step employed for all the cases, which results in a variable CFL number, greater for the $\mathrm{Ma}_\infty=1.3$ case. This behavior is common to several numerical methods, where an optimal trade off between accuracy, stability and speed convergence has to be found in terms of CFL number.
\begin{figure}[!t]
\centering
\includegraphics[width=1.0\columnwidth]{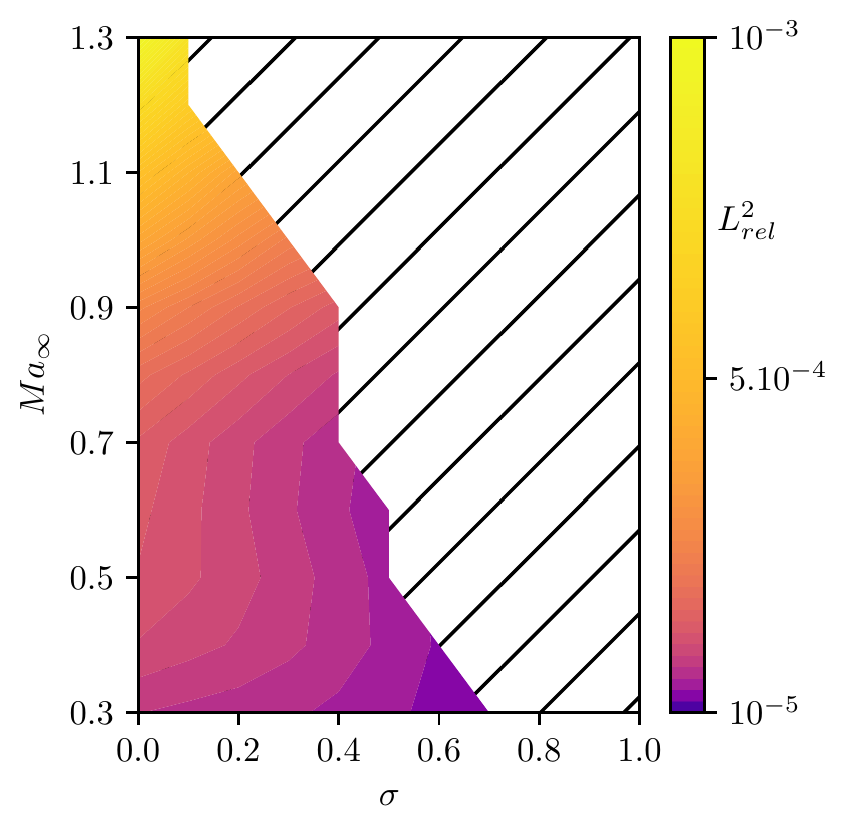}
\caption{Color map of the $L^2$ relative error on pressure after 50 convective times as a function of the Mach number and $\sigma$ parameter. Simulations have been performed in intervals of $0.1$ for both $\sigma$ and $\mathrm{Ma}$, representing in total hundred simulations. The dashed region refers to where the computation diverged or the total energy maximum was greater at the end of the computation.}
\label{fig: L2err covo map}
\end{figure}
Fig.\ref{fig: L2err covo map} shows a sensitivity analysis on the $L^2$ relative error depending on the $\sigma$ parameter. For different Mach number regimes, one can observe that the accuracy and stability of the solution depends on the parameter $\sigma$ similarly to the original model~\cite{Feng2019}. The higher the Mach number is, the smaller the $\sigma$ parameter has to be, decreasing the accuracy of the solution for the benefit of stability. Thus, the more robust choice of parameters is $\sigma=0$ independent of the Mach number, allowing for the model to be valid from subsonic to supersonic regime. This value will be chosen in the following to compare the present enhanced model against the original model~\cite{Feng2019}.
\begin{figure}[!t]
\centering
\includegraphics[width=1.0\columnwidth]{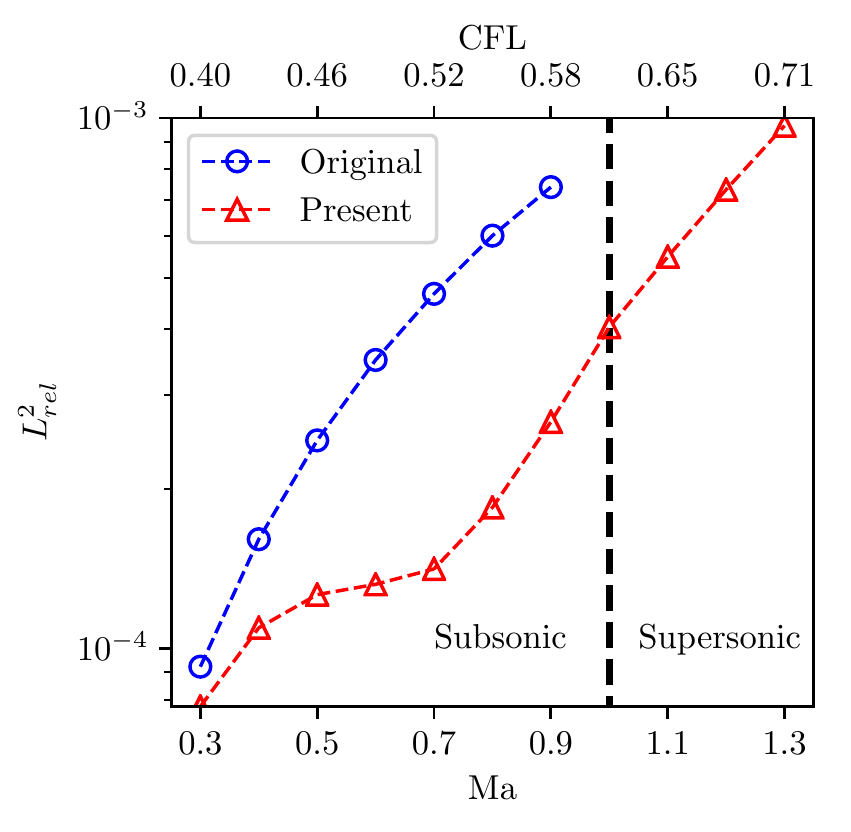}
\caption{$L^2$ relative error on pressure after 50 convective times in function of the Mach number (and equivalent CFL number). Each symbol correspond to a simulation. The lack of points means that the computation diverged or the total energy maximum was greater at the end of the computation. The original model~\cite{Feng2019} appears to be restricted to subsonic flows whereas the present one allows supersonic regimes.}
\label{fig: L2err covo}
\end{figure}

\begin{figure}[!t]
\centering
\includegraphics[width=1.0\columnwidth]{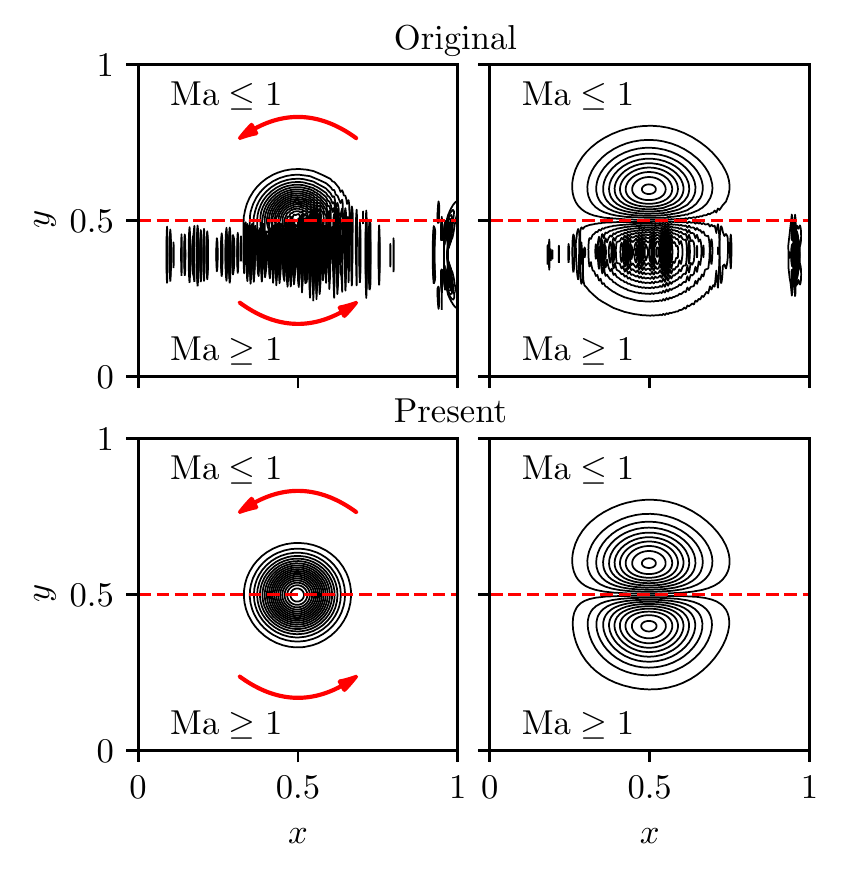}
\caption{Convected isentropic vortex after $3$ convective times at $\mathrm{Ma}_\infty=1$ and where the two, subsonic and supersonic, regions are distinguished by the dashed line. Iso-contour of density (left) and $x$ component of the velocity (right) are displayed. Arrows indicate the sens of rotation of the vortex. The original model~\cite{Feng2019} model (top), shows instabilities in the supersonic region while the present model (bottom), is stable and present excellent results in terms of isotropy}
\label{fig: ReviewIso_Rho_VX_3tr}
\end{figure}

Fig.~\ref{fig: L2err covo} shows the $L^2$ relative error of the two models after $50$ convective times of vortex. Different Mach numbers are considered from subsonic to supersonic regime. For all the advection cases, the present model shows to be more accurate than the original of which, the latter, turns out to be unstable for Mach numbers higher or equal to one. For the specific case of $\mathrm{Ma}=1$, Fig.~\ref{fig: ReviewIso_Rho_VX_3tr} shows the iso contours of density and velocity for the two models, just before the computation diverge for the original model. In contrary to the present model, it can be observe that, the original one, produces instabilities located in the supersonic region and not in the subsonic one. This clearly confirms observations made from Fig~\ref{fig: L2err covo}, where no stable computations using the original model has been found for supersonic cases, restricting its application range to subsonic regimes.

Throughout the study of this test case, it has been shown that the present HLBM model permits to increase the stability domain of the original one in terms of Mach number, allowing now the computation of supersonic flows.

\subsection{\label{sec:Acoustic damping}Acoustic damping}
The effect of the correction term $E_{2,\alpha\beta}$ in Eq.~(\ref{eq: E2 corr term}) of the present model, is assessed studying the decay of a well  resolved sinusoidal acoustic wave. A quasi one dimensional periodic domain of $L_x=1.10^{-1}\,\mathrm{m}$, discretized on $[200\times2]$ points is considered. 
The acoustic pressure profile reads:
\begin{equation}
\widetilde{p}(x)= \Delta p \sin \left( \frac{2\pi}{L_x} x \right),
\label{eq: Pressure_acoustic wave}
\end{equation}
with $\Delta p = 5\,\mathrm{Pa}$ being the amplitude of the perturbation.
The present acoustic is considered isentropic, thus from the Laplace laws, the temperature field is found to be:
\begin{equation}
\widetilde{T}(r)= \left( \frac{\widetilde{p}(r)}{p_{\infty}} \right)^{\frac{\gamma_g-1}{\gamma_g}}  T_{\infty},
\label{eq: Temperature_acoustic}
\end{equation}
and the velocity field is given by:
\begin{equation}
\widetilde{u_x}= \frac{\widetilde{p}}{\rho c_\infty}.
\label{eq: Init u acoustic wave}
\end{equation}
Finally the initial field reads:
\begin{equation}
p=p_\infty+\widetilde{p}  ~ ; ~ T = T_\infty+\widetilde{T} ~; ~  u_x=\widetilde{u_x},
\label{eq: Init_acoustic}
\end{equation}
where $p_\infty=101325\,\mathrm{Pa}$, $T_\infty=300\,\mathrm{K}$.
In this case, $T_r=1478.75\,\mathrm{K}$, $\sigma=0.95$, a Prandtl number of $\mathrm{Pr}=0.71$ is considered and the dynamic viscosity is taken as $\mu=0.01 \,\mathrm{kg.m^{-1}.s}^{-1}$. The expected pressure decay rate reads~\cite{wissocq2019thesis}:
\begin{equation}
\max\left(p\right) = p_\infty + \Delta p \exp \left( - \alpha \left(\frac{2\pi}{L_x}\right)^2 t \right),
\label{eq: Pressure_acoustic wave decay}
\end{equation}
with
\begin{equation}
\alpha= \frac{ D-1}{D} \nu + \frac{\gamma_g-1}{2}\frac{\nu}{\mathrm{Pr}},
\label{eq: coef acoustic wave decay}
\end{equation}
the acoustic decay rate for a fluid with zero bulk viscosity. This decay rate is computed from the simulation using the similar process described in Sec.~\ref{sec:shear wave}.
\begin{figure}[!t]
\centering
\includegraphics[width=1.0\columnwidth]{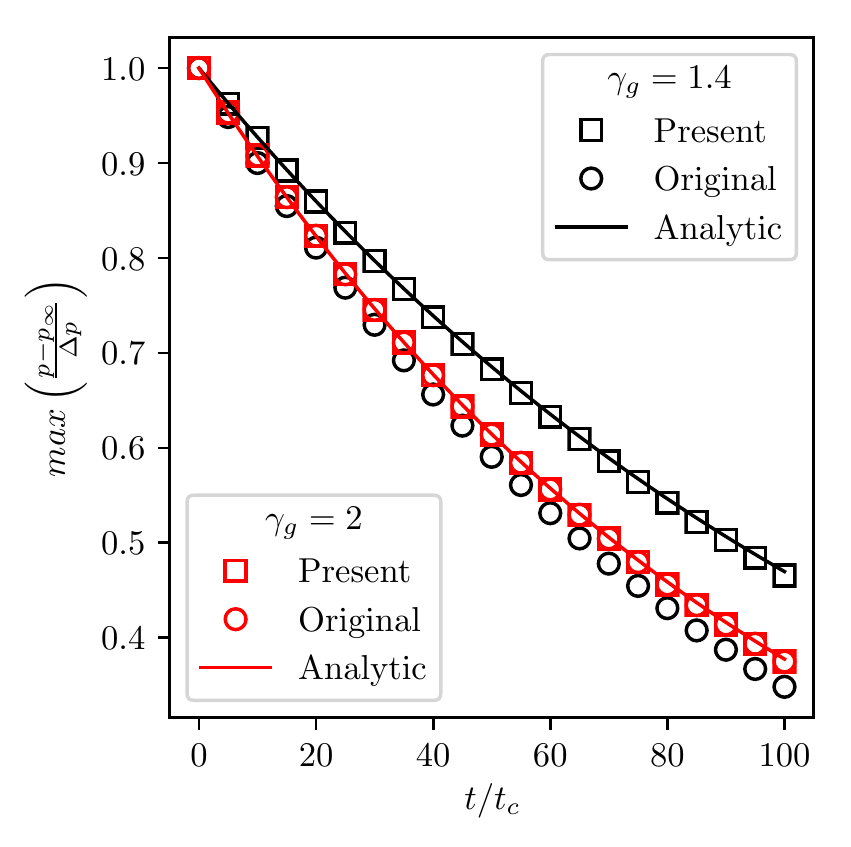}
\caption{Acoustic wave decay over dimensionless time $t/t_c$ with $t_c=L_x/\sqrt{\gamma_g r T_\infty}$. As expected, a correct acoustic dissipation for monatomic value of the heat capacity ratio is retrieved for both models. For the polyatomic case, the original model~\cite{Feng2019} deviates from the analytic while the correct dissipation is retrieved by the present one.}
\label{fig: Acoust decay}
\end{figure}
If the error is not corrected for a freely chosen value of $\gamma_g$, i.e. $\gamma_g=1.4$ in the present case, a spurious bulk viscosity appears, increasing the acoustic decay rate.
The results presented on Fig.\ref{fig: Acoust decay}, clearly show that the correct acoustic decay rate is retrieved by the present model (square symbols), compared to the analytic (solid lines), and this for monatomic and polyatomic values of the heat capacity ratio. However, the original model~\cite{Feng2019} (circles symbols) over dissipate the acoustic wave in the polyatomic case, due to a spurious bulk viscosity.
For a monatomic value $\gamma_g=\gamma_{lb}$, no error is found for both original and present models, which confirms the statement in Sec.\ref{subsec: Corr terms enhancement}.

\subsection{\label{sec:couette}Thermal Couette flow}

Here the viscous heat production and dissipation of the scheme is assessed using the thermal Couette flow test case. 
It consists of two parallel plates, one stationary, and one translating along the $x$ direction at constant velocity $U=Ma*c_\infty$. This motion generates a shear flow with a linear velocity profile, which produces a temperature increase by viscous friction. The steady state is reached once the heat dissipation balance the heat production. The analytical solution for the temperature profile between the plates exists in two variants consisting by, either keeping the plates at different temperature $\Delta T\neq0$, or at the same temperature $\Delta T=0$. In the first case, the analytical solution only depends on the Eckert and the Prandtl numbers while in the second case it depends on the heat capacity ratio, the Mach and Prandtl numbers. This last analytical solution reads:
\begin{equation}
\frac{T - T_w}{T_w} = \frac{y}{H} \mathrm{Pr} \mathrm{Ma}^2 \frac{\gamma_g-1}{2} \left( 1 -\frac{y}{H} \right), 
\label{eq: couette analy}
\end{equation}
with $T_w$ the temperature of the plate and $H$ the distance between the two plates.
A simulation domain of $\left[2\times100\right]$ points was considered and regularized boundary conditions described in~\cite{Feng2019} were adopted. $T_{r}=1500\,\mathrm{K}$ and $\sigma=0.9$ so that 90\% of the viscous stress tensor is modeled by projection of the $f_i$ and 10\% by finite difference. The temperature of the plates is taken at $T_w=T_\infty=300\,\mathrm{K}$ and the initial pressure of the fluid is $p_\infty=101325\,\mathrm{Pa}$.  
\begin{figure}[!t]
\centering
\includegraphics[width=1.0\columnwidth]{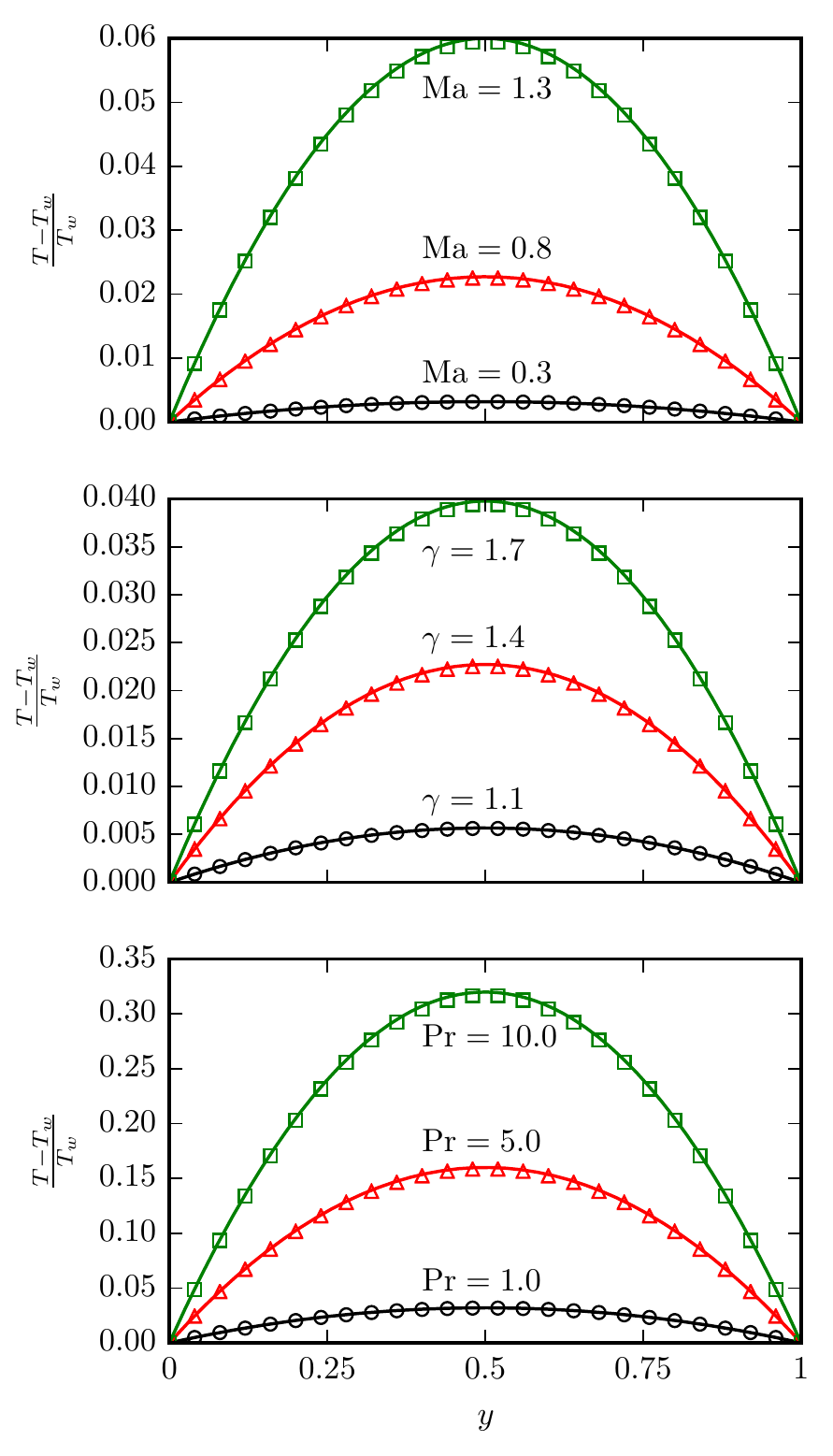}
\caption{Steady state temperature profile along the $y$ direction of the thermal Couette flow, for different Mach numbers, Prandtl numbers and heat capacity ratios. Symbols correspond to HLBM and solid lines to the analytic. For all the plots, the missing variables are taken to $\mathrm{Ma}=0.8$, $\mathrm{Pr}=0.71$ and $\gamma_g=1.4$.}
\label{fig: Couette}
\end{figure}
Fig.~\ref{fig: Couette} shows the temperature profiles obtained by the present model (symbols), for different Mach numbers, heat capacity ratios and Prandtl numbers, and are compared to the analytical solution (solid lines). For all the cases, the results are in good agreement with the analytical solution. Thus, the new discretization of viscous heat production terms Eq.~(\ref{eq:viscous heat discret}), partially computed with the LBM part through $a^1_{\alpha\beta}$ and finite differences, shows to be a potential substitute to a full finite difference discretization. Finally, it is worth noting that for this test case, the major source of error is induced by the boundary conditions, and the correction term has no role to play in this configuration due to the nature of the flow (unidirectional and uniform).

\section{\label{sec:Numerical validation}Numerical validation and model capabilities}

In the previous section, the limitation to subsonic flows of the original model~\cite{Feng2019} has been demonstrated throughout different test cases. 
The improvements of the present model permits to remove this limitation, and have been successfully validated on shear diffusion, vortical convection, acoustic decay, and viscous heat production and dissipation cases. These test cases have been done from subsonic to supersonic regimes, and comparisons with the original model~\cite{Feng2019}, ensured no regressions for subsonic flows.  
Now the aim is to perform computations at high Mach cases with discontinuities using the present HLBM scheme. However in a first time, and for the sake of completeness with regard to the previous test cases, the acoustic will be assessed in terms of propagation. Indeed, an isentropic speed of sound must be retrieved, which will finish to validate all the physical phenomena of interest related to the Navier-Stokes Fourier equations. In a second time, a shock tube test will be computed for a relatively low viscosity value, showing the robustness of the method to treat flows discontinuities. Eventually, a shock-vortex interaction case is performed, permitting the assessment of the present model in a configuration where all the physical phenomena previously validated are combined.

\subsection{\label{sec:Acousticcases}Acoustic cases}

In order to assess the acoustic propagation, two test cases are considered.
The first one is a Gaussian shaped acoustic pulse in a mean flow at two different Mach numbers, respectively $\mathrm{Ma}=0$ and $\mathrm{Ma}=1$ . The small acoustic perturbation reads:
\begin{equation}
\widetilde{p}(r)= \Delta p \exp{\left(-\frac{r^2}{2}\right)},
\label{eq: Pressure_acoustic}
\end{equation}
with $\Delta p = 10 \,\mathrm{Pa}$ its amplitude. The acoustic is considered isentropic, thus the temperature field can be found by the Laplace laws from Eq.~(\ref{eq: Temperature_acoustic}).
Finally the initial field reads:
\begin{equation}
p=p_\infty+\widetilde{p}  ~ ; ~ T = T_\infty+\widetilde{T} ~; ~  u_x=\mathrm{Ma}\,c_\infty  ~;~ u_y=0.
\label{eq: Init_acoustic}
\end{equation}
Six points have been taken in the pulse radius $R$, $T_\infty=300\,\mathrm{K}$ and $p_\infty=101323\,\mathrm{Pa}$. The numerical domain is a one meter square box discretized on $[201\times201]$ points. In the null Mach case, $\sigma=0.95$ is considered and $\sigma=0$ is employed in the sonic case. 
The reference temperature is taken to $T_{r}=1478.75\,\mathrm{K}$ for both cases. The dynamic viscosity is set to $\mu=1.10^{-5} \,\mathrm{kg.m^{-1}s}^{-1}$, the heat capacity ratio to $\gamma_g=1.4$ and the Prandtl number to $\mathrm{Pr}=0.71$.  The $\mathrm{Ma}=0$ pulse is initialized in the center of the numerical domain $[0.5;0.5]$ while the $\mathrm{Ma}=1$ pulse is placed at $[0.1;0.5]$. For both cases, the simulation lasts $t = 0.4t_c$ with $t_c=L/c_\infty$ with $L$ the domain length. For this particular choice of time and initial position, both solutions are expected to be the equal. 
\begin{figure}[!t]
\centering
\includegraphics[width=1.0\columnwidth]{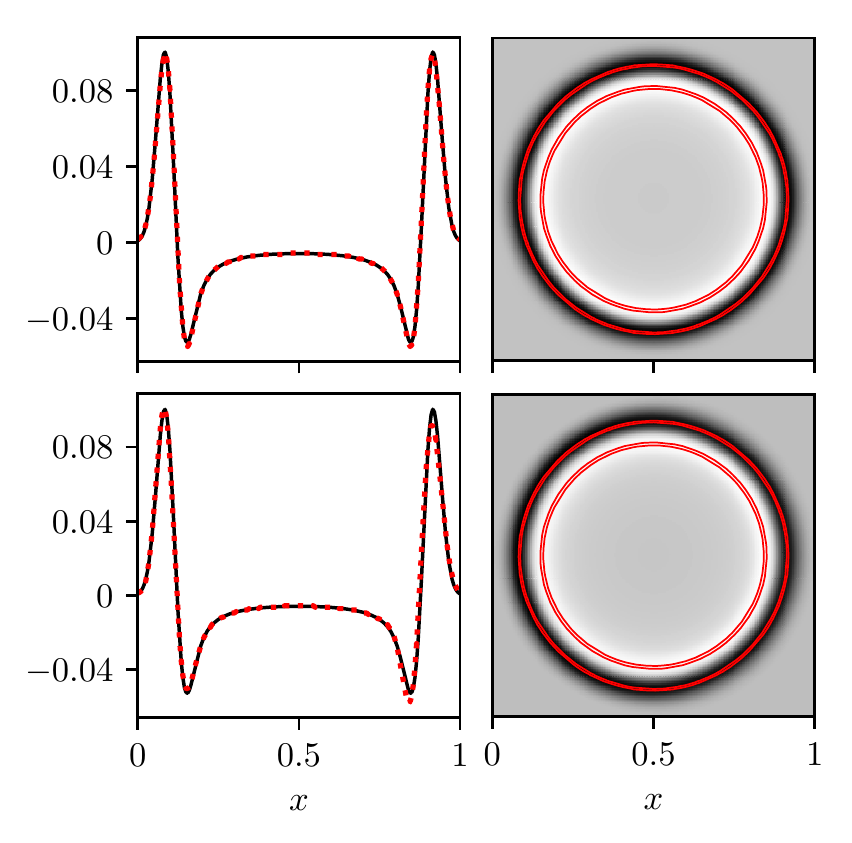}
\caption{Pressure profile for $y=0.5$ (left) and pressure color map (right) of an acoustic pulse simulation for $\mathrm{Ma}=0$ (top) and $\mathrm{Ma}=1$ (bottom). On the left graphs, the analytical solution corresponds to the solid black line and the dotted red curve represents the observed response. On the right graphs, the two red contours correspond to the min and max value of the analytical solution.}
\label{fig: Pulse}
\end{figure}
Fig.~\ref{fig: Pulse} shows good agreement with the analytical solution~\cite{tam1993dispersion} for the subsonic and sonic cases. However, small dispersion and dissipation defects appear on the downstream acoustic part for $\mathrm{Ma}=1$, meaning that the numerical error grows with respect to the $\mathrm{CFL}$ number. Nonetheless, this trend is in agreement with the observations in Sec.~\ref{sec:covo}. The isocontours show correct isotropy of the solution in both cases, meaning there is no flagrant directional bias in the acoustic propagation.

Next, the isentropic speed of sound is validated for different heat capacity ratios and fluid temperatures. To that end, the velocity of a well resolved  Gaussian shaped acoustic wave, with 20 points in the standard deviation $R$, is measured. The fluid is considered at rest and the pressure, temperature and velocity profiles are defined according to Eq.~(\ref{eq: Pressure_acoustic}), Eq.~(\ref{eq: Temperature_acoustic}) and Eq.~(\ref{eq: Init u acoustic wave}) respectively.
The speed measurement is done after 10 rounds of a one meter periodic domain discretized by $[200\times2]$ points. The time step remains the same as before and $\sigma=0.95$ has been used.
\begin{figure}[!t]
\centering
\includegraphics[width=1.0\columnwidth]{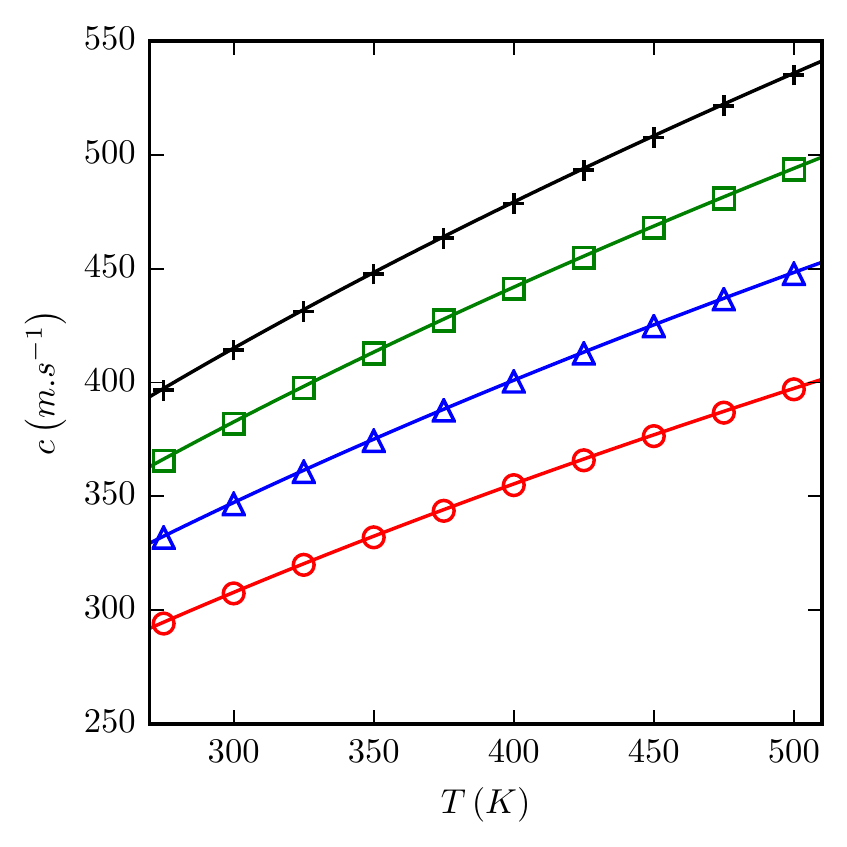}
\caption{Sound speed measurement for different fluid temperature and heat capacity ratio. Each symbols correspond to a simulation. Here \protect\RedCircle corresponds to $\gamma_g=1.1$~;~\protect\BlueTriangle to $\gamma_g=1.4$~;~\protect\GreenSquare to $\gamma_g=1.7$ and \protect\BlackCross to $\gamma_g=2.0$. Solid lines corresponds to the theoretical values.}
\label{fig: Ac_wave}
\end{figure}
Fig.~\ref{fig: Ac_wave} shows good agreements between the computed velocity (symbols) and the theoretical value (solid lines). Thus it validates the perfect gas coupling Eq.~(\ref{eq: perfect gas}) and the isentropic speed of sound Eq.~(\ref{eq: speed of sound}) of this HLBM scheme.

\subsection{\label{sec: Sod}Shock tube problem}
Here the present model is assessed on a one-dimensional flow including discontinuity.
The flow of the shock tube problem includes a shock wave,  a slip line (contact surface) and an expansion wave permitting to assess a numerical scheme for these three physical phenomena. This test case is performed with $\gamma_g=1.4$, $\mu=1.10^{-5}\,\mathrm{kg.m^{-1}.s}^{-1}$, $\sigma=0.4$ and $T_r=1460\,\mathrm{K}$. The initial conditions read:
\begin{equation}
(\rho, u, p)=
\begin{cases}
(3\rho_\infty,0,3p_\infty),&0\leq x<0.5\\
(\rho_\infty,0,p_\infty),&0.5\leq x\leq1
\end{cases},
\end{equation}
where $T_\infty=300\,\mathrm{K}$, $p_\infty=101325 \,\mathrm{Pa}$ and $\rho_\infty = p_\infty/\left(rT_\infty\right)$. We chose a quasi one dimensional domain of one meter with $401\times2$ points.
\begin{figure}[!t]
\centering
\includegraphics[width=1.0\columnwidth]{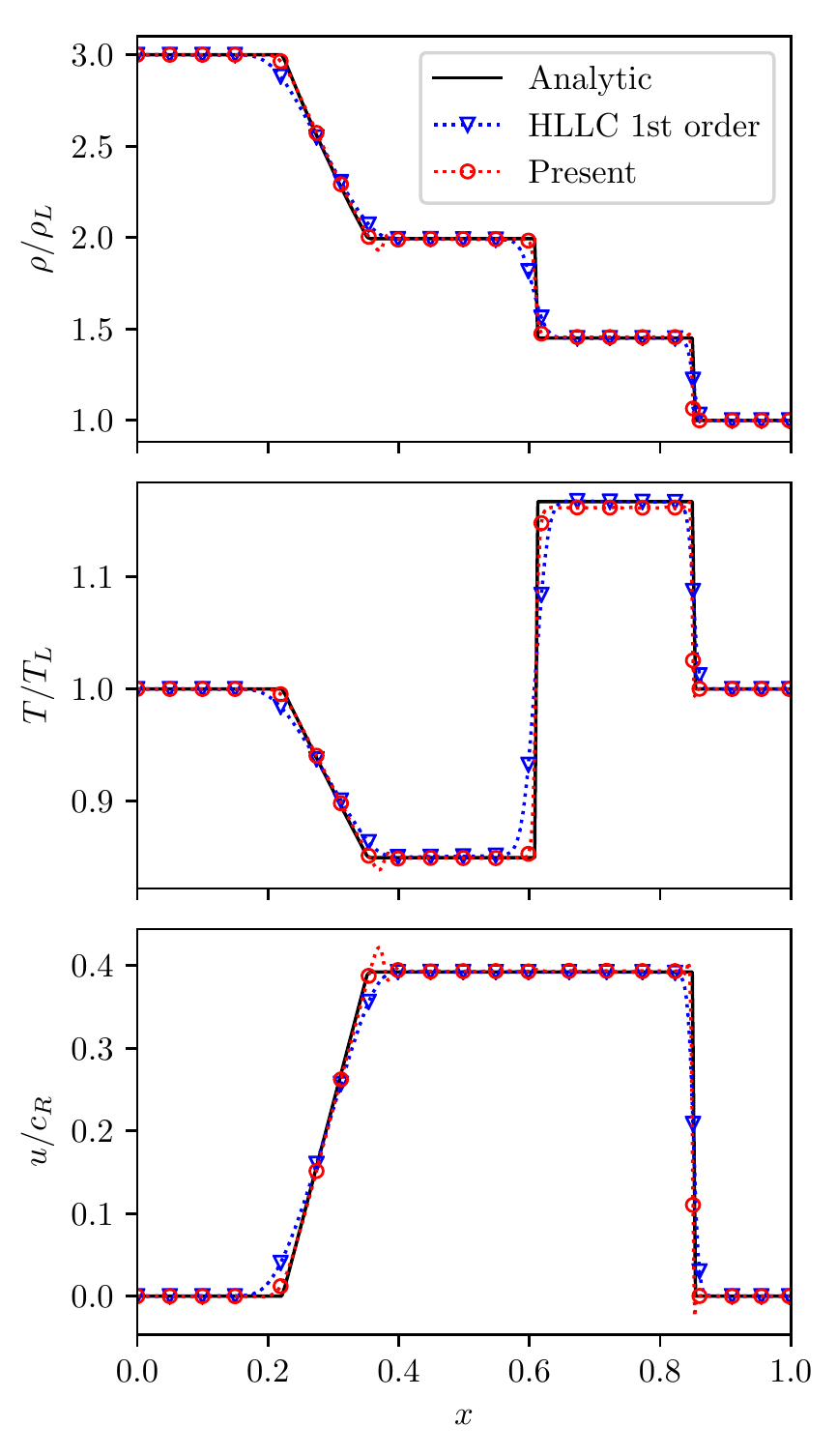}
\caption{Sod shock tube problem. The present model is compared to a first order HLLC Riemann solver~\cite{toro2013riemann} and to the analytical solutions.}
\label{fig: Sod}
\end{figure}

Fig.~\ref{fig: Sod} shows on different profiles the overall agreement of the computed solutions (circle symbols) compared to the analytic (solid lines) and to a first order HLLC (Harten, Lax and van Leer plus Contact) Riemann solver~\cite{toro2013riemann} (triangle symbols). Nonetheless, a spurious oscillation at the bottom of the rarefaction wave is present which bias the solution. However, this defect can be easily handled  since it disappears for a larger choice of viscosity. This viscosity can be added locally on the discontinuity with the help of a shock sensor in the manner of~\cite{coreixas2018high}. Moreover, the temperature profile between the shock and the contact discontinuity is slightly underestimated. Indeed this is due to the non conservative nature of the energy equation, and it is well known that this form does not permit to capture efficiently the flow discontinuities within a numerical framework~\cite{Toro1965,toro2013riemann}. To overcome this loss, Nie $et$ $al.$ in~\cite{Nie2009}, used an artificial production term in the energy equation, benched on several shock tube cases, to balance this defect. But this leads to an additional free parameter dependencies of the model which is not the purpose in this work. Although this defect increases for higher ratios between the left and the right states, in the present case the model gives satisfying results, representative of a typical shock encountered in transonic aeronautical flows \cite{Nie2009}.

\subsection{\label{sec: Shock vortex}Shock vortex interaction}

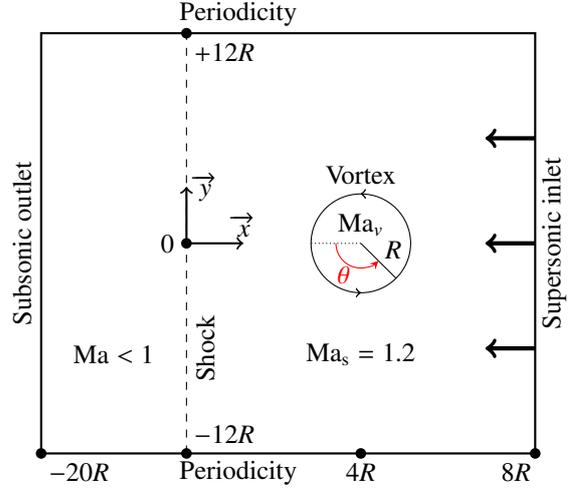
\begin{figure}[!t]
\centering
\begin{tikzpicture}
%Domain definition
\tikzmath{
%Domain
\Lx = 6.5; % Domain length X
\Ly = 0.8571428*\Lx;   % Domain length Y 
%Vortex
\R = \Ly/8.5; % Vortex radius 
\D = 3.5*\R; %Vortex distance from Bnd
\Rx = \Lx - \D; %Vortex x position
\Ry = \Ly/2; % Vortex y position
\RvRx = \Rx + \R/2 +0.1; % R letter position 
\RvRy = \Ry+0.15; % R letter position 
\RvThetax = \Rx - \R/2 +0.1; % R letter position 
\RvThetay = \Ry-0.18; % R letter position 
%Shock
\Sx1=\Rx- \D;
\Sy1=0;
\Sx2=\Sx1;
\Sy2=\Ly;
% Space for anotations
\dd=0.25;
% Axis arrow length
\Axs=0.75;
}
\coordinate (A) at (0,0);
\coordinate (B) at (\Lx,0);
\coordinate (C) at (\Lx,\Ly);
\coordinate (D) at (0,\Ly);
\coordinate (Rv) at (\Rx,\Ry); %Vortex center
\path (Rv) + (-\R,0) coordinate (Rvs); % End of the radius line dotted of the vortex
\path (Rv) + (0.7071*\R,-0.7071*\R) coordinate (Rvs2); % Radius line of the vortex
\path (Rv) + (-\R/2,0) coordinate (Rvs3); % starting point of the theta angle
\path (Rv) + (0.7071*\R/2,-0.7071*\R/2) coordinate (Rvs4); % ending point of the theta angle
\coordinate (RvR) at (\RvRx,\RvRy); % R letter position
\coordinate (Rvtheta) at (\RvThetax,\RvThetay); % R letter position
% Draw the domain
\draw [thick] (A) -- (B) -- (C) -- (D) -- (A);
\draw (-\dd,\Ly/2) node[rotate=90]{Subsonic outlet};
\draw (\Lx+\dd,\Ly/2) node[rotate=90]{Supersonic inlet};
\draw (\Lx/2-\R,\Ly+\dd) node[rotate=0]{Periodicity};
\draw (\Lx/2-\R,-\dd) node[rotate=0]{Periodicity};
% Draw the vortex
\draw (Rv) -- (Rvs2);
\draw [densely dotted] (Rv) -- (Rvs);
\draw[->,red,>=stealth] (Rvs3) arc (0:135:-\R/2);%angle theta. -\R provide the other curvature side !!!
\draw (Rvtheta) node[below]{$\textcolor{red}{\theta}$};
\draw [decoration={markings, mark=at position 0.25 with {\arrow{>}}},postaction={decorate}] [decoration={markings, mark=at position 0.75 with {\arrow{>}}},postaction={decorate}] (Rv) circle  (\R);
\draw (RvR) node[below]{$R$};
\draw (\Rx,\Ry+\R+\dd) node[rotate=0]{Vortex};
\draw (\Rx,\Ry+\dd) node[rotate=0]{$\mathrm{Ma}_v$};
% Draw the Shock
\coordinate (S1) at (\Sx1,\Sy1);
\coordinate (S2) at (\Sx2,\Sy2);
\draw [dashed] (S1) -- (S2);
\draw (\Sx1+\dd,\Ly*1/4) node[rotate=90]{Shock};
% Draw the tick and axis
\draw [thick] [->] (\Sx1,\Ly/2) -- (\Sx1+\Axs,\Ly/2); % X axis
\draw (\Sx1+\Axs,\Ly/2+\dd) node[rotate=0]{$\overrightarrow{x}$};
\draw [thick] [->] (\Sx1,\Ly/2) -- (\Sx1,\Ly/2+\Axs); % Y axis
\draw (\Sx1+\dd,\Ly/2+\Axs) node[rotate=0]{$\overrightarrow{y}$};
\draw (\Sx1-\dd,\Ly/2) node[rotate=0]{$0$};\fill (\Sx1,\Ly/2) circle[radius=2pt];
\draw (\Rx,-\dd) node[rotate=0]{$4R$};\fill (\Rx,0)  circle[radius=2pt];
\draw (\Lx-\dd,-\dd) node[rotate=0]{$8R$};\fill (\Lx,0)  circle[radius=2pt];
\draw (2*\dd,-\dd) node[rotate=0]{$-20R$};\fill (0,0)  circle[radius=2pt];
\draw (\Sx1+2*\dd,\dd) node[rotate=0]{$-12R$};\fill (\Sx1,0)  circle[radius=2pt];
\draw (\Sx1+2*\dd,\Sy2-\dd) node[rotate=0]{$+12R$};\fill (\Sx1,\Sy2)  circle[radius=2pt];
%Miscellaneous
\draw (\Rx,2*\R) node[rotate=0]{$\mathrm{Ma_s}=1.2$};
\draw (\Sx1/2,2*\R) node[rotate=0]{$\mathrm{Ma}<1$};
% Draw the arrow for the flow
\draw [ultra thick] [->] (\Lx,\Ly/4) -- (\Lx-\R,\Ly/4); 
\draw [ultra thick] [->] (\Lx,2*\Ly/4) -- (\Lx-\R,2*\Ly/4); 
\draw [ultra thick] [->] (\Lx,3*\Ly/4) -- (\Lx-\R,3*\Ly/4); 

\end{tikzpicture}

\caption{Numerical setting for the shock vortex cases. For all the cases, the Mach number of the shock wave is taken at $\mathrm{Ma}_s=1.2$. The vortex has an anticlockwise circulation and propagates from the right to the left of the domain.}
\label{Shock_Vortex_Layout}
\end{figure}

Here the present model is finally assessed for a two dimensional unsteady compressible viscous test case. 
It consists in the interaction between a vortex and a planar stationary shock wave, at advection Mach number of $\mathrm{Ma}_s=1.2$.
The results are compared with the direct numerical simulation (DNS) of~\cite{inoue1999sound}.
A computational domain of $[-20R,+8R]\times[-12R,+12R]$ on a $1681\times1441$ mesh is adopted, with R the vortex radius taken at $R=1\,\mathrm{m}$.
The planar shock wave is specified at $x=0$ by imposing the density, velocity and pressure corresponding to the left and right states of the steady shock, using the Rankine-Hugoniot relations. The freestream (pre shock $x>0$) quantities are taken to be $p_\infty=101325\,\mathrm{Pa}$ and $T_\infty=300\,\mathrm{K}$ while the time step is defined with $T_r=1404\,\mathrm{K}$.
The vortex is initialized far enough of the shock at $x=4R$ to avoid interaction issues and travels from the right to the left of the shock. Its velocity, density and pressure profiles are defined according to Eq.~(\ref{eq: V covo}), Eq.~(\ref{eq: rho covo}) and Eq.~(\ref{eq: p covo}) respectively. The characteristic time is defined as $t_c=R/c_\infty$ where $c_\infty$ is the speed of sound before the shock. It is worth noting that in order to compare the results with~\cite{inoue1999sound}, $t=0\times t_c$ corresponds to where the vortex is positioned at $x=2R$ from the shock. Thus in the current case, the vortex already travels the distance $2R$ when $t=0\times t_c$ is reached.
The boundary conditions are imposed similarly to~\cite{Feng2019} and their type and the layout of the test case can be found in Fig.~\ref{Shock_Vortex_Layout}. Case A with $\mathrm{Ma}_v=0.5$ and $\mathrm{Re}=400$ and case C with $\mathrm{Ma}_v=0.25$ and $\mathrm{Re}=800$ from ref.~\cite{inoue1999sound} have been performed. 

\begin{figure}[!t]
\centering
\includegraphics[width=0.98\columnwidth]{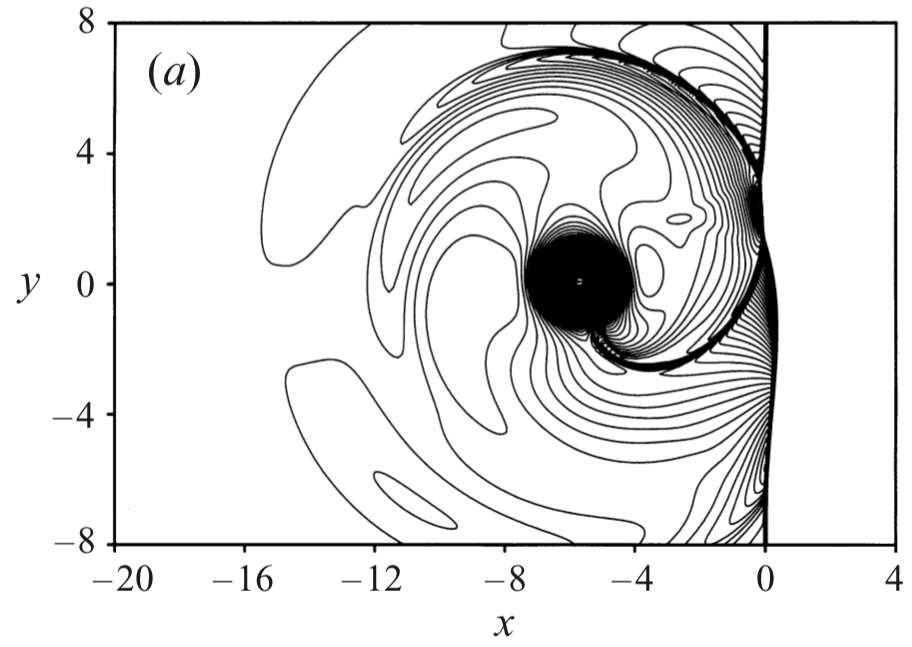}
\includegraphics[width=1.0\columnwidth]{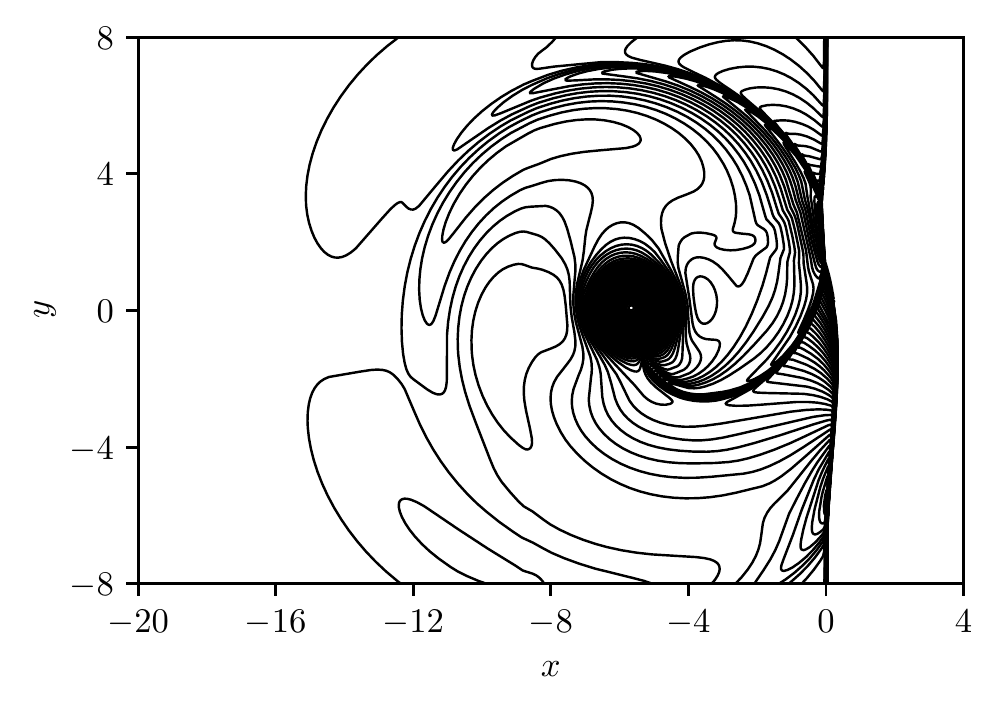}
\caption{Case A with $\sigma=0$. Density isocontour, ref.~\cite{inoue1999sound} (top) and present HLBM (bottom). Contour ranges from $\rho_{min}=0.92$ to $\rho_{max}=1.55$ with an increment of 0.0053.}
\label{fig:rho_Case_A}
\end{figure}
For case A, the isocontours of density on Fig.~\ref{fig:rho_Case_A}, clearly show the deformation of the shock wave, from which, a pair of reflected shock waves emanate due to the interaction with the vortex. The result is qualitatively in good agreements with the DNS isocontours, meaning that the overall feature of the flow is well modeled by the present scheme. 
\begin{figure}[!t]
\centering
\includegraphics[width=1.0\columnwidth]{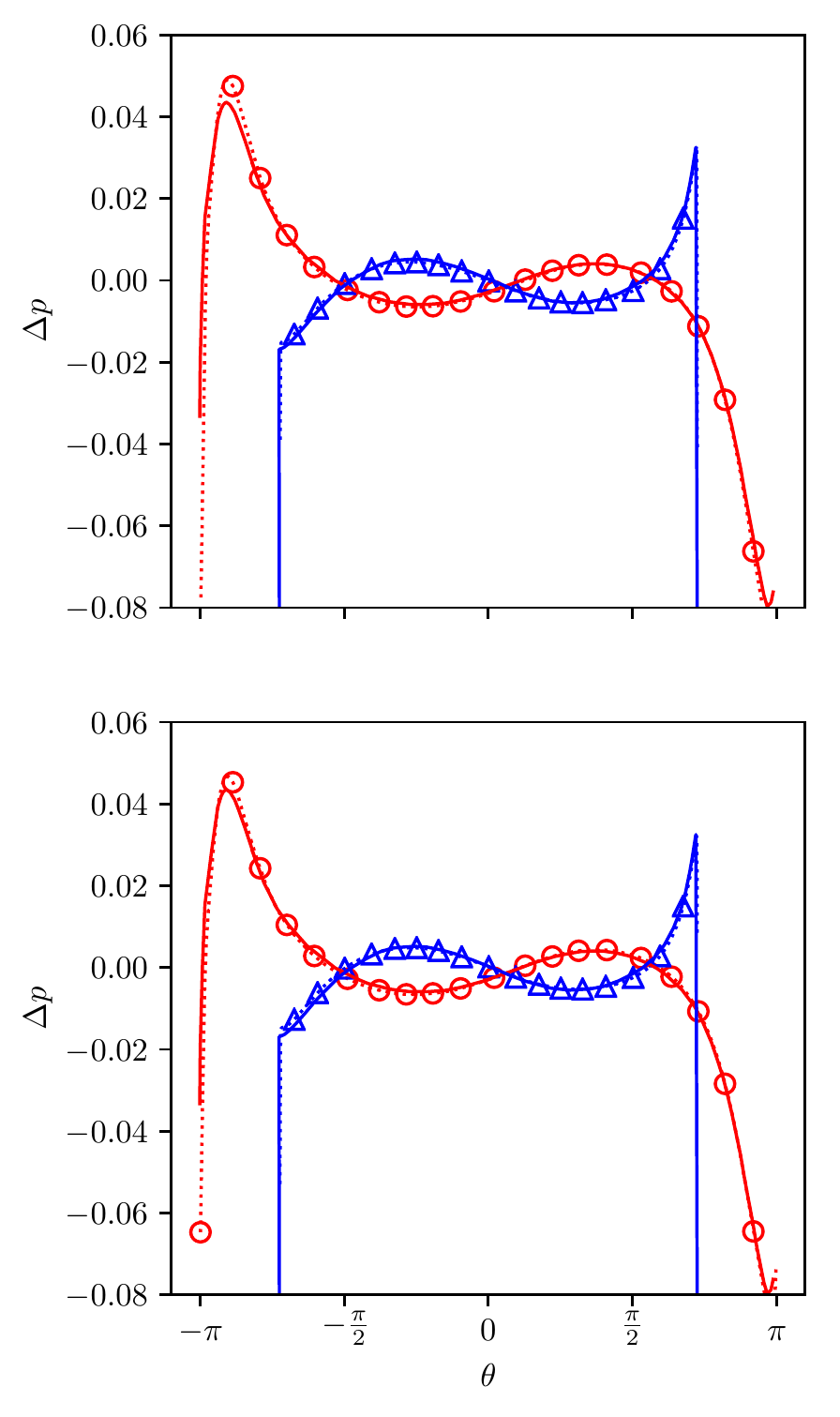}
\caption{Case C with $\sigma=0$ (top) and $\sigma=0.9$ (bottom). Circumferential cut of the pressure variation $\Delta p$ at $t_c=6$. Here symbols correspond to the present HLBM model while solid line to the DNS~\cite{inoue1999sound}. 
The symbols \protect\BlueDottedTriangle correspond to the precursor at $r=6$ while \protect\RedDottedCircle refer to the second sound at $r=3.7$.}
\label{fig: DP_Case_C_Circumferencial}
\end{figure}

\begin{figure}[!t]
\centering
\includegraphics[width=1.0\columnwidth]{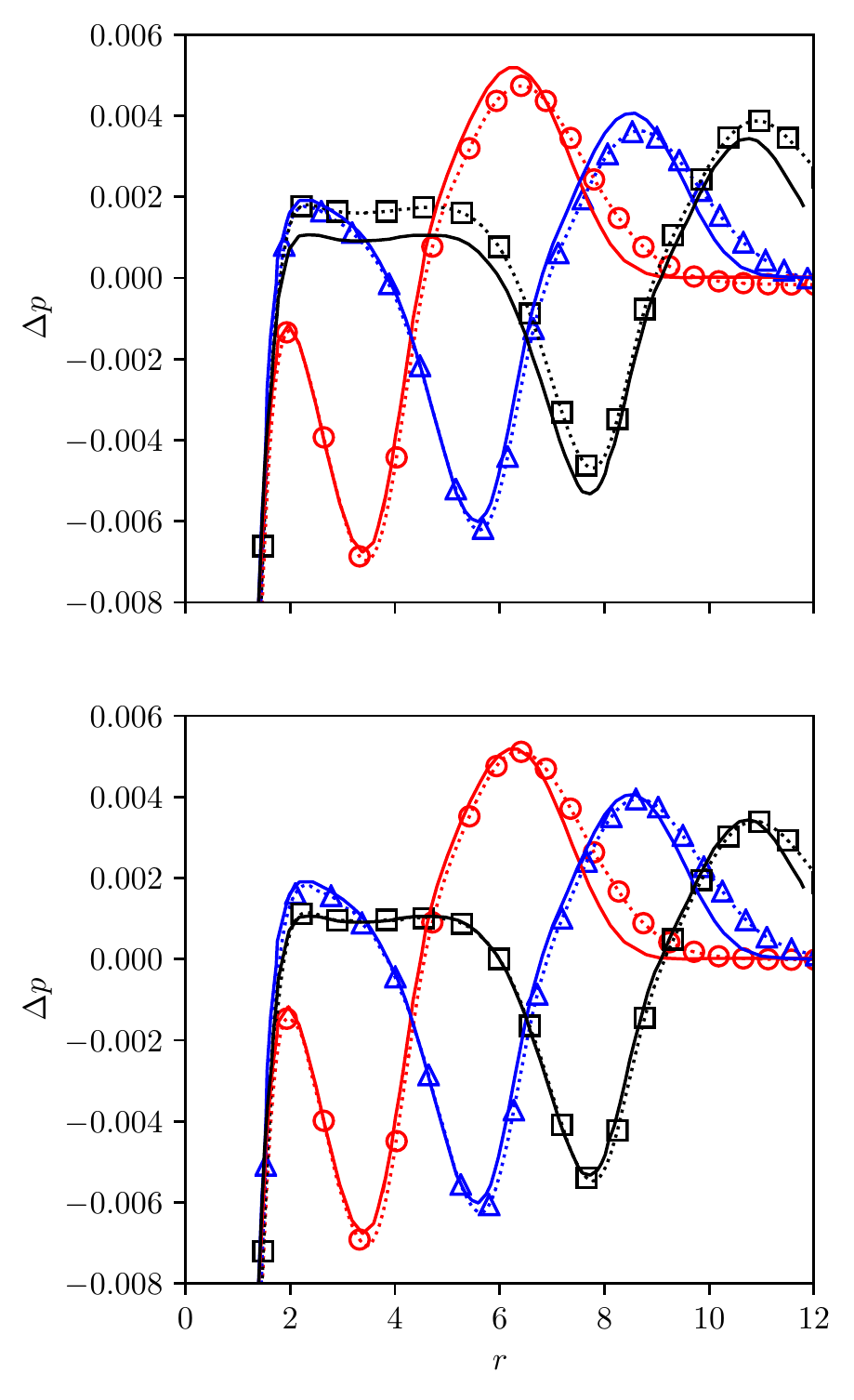}
\caption{Case C with $\sigma=0$ (top) and $\sigma=0.9$ (bottom). Radial cut at $\theta=-\pi/4$ of the pressure variation $\Delta p$. Solid lines correspond to the DNS~\cite{inoue1999sound} and symbols refers to the present HLBM for $t_c=6$ \protect\RedDottedCircle, $t_c=8$ \protect\BlueDottedTriangle and $t_c=10$ \protect\BlackDottedSquare }
\label{fig: DP_Case_C_Radial}
\end{figure}
For quantitative results, the acoustic sound generation of case C is compared to the DNS~\cite{inoue1999sound}. This sound is defined as:
\begin{equation}
\Delta p = \frac{p-p_s}{p_s},
\end{equation}
with $p_s$ the pressure behind the shock wave. Fig.~\ref{fig: DP_Case_C_Circumferencial} and Fig.~\ref{fig: DP_Case_C_Radial}, compare the circumferential and radial distributions of the sound pressure $\Delta p$ computed by the present model (symbols) to the DNS results (solid lines). This is done for two different values of $\sigma$, $\sigma=0$ and $\sigma=0.9$. As shown in the previous section~\ref{sec:covo}, in low viscosity cases a small value of $\sigma$ is mandatory to obtain stable results which decreases the accuracy of the solution. However, the present case is viscous enough to get a stable simulation even with high value of $\sigma$ permitting the study of its impact on the solution.
When the vortex interacts with the shock, a precursor wave appears first, followed by a second acoustic wave (the second sound)~\cite{inoue1999sound}, which are both of quadrupolar nature. Fig.~\ref{fig: DP_Case_C_Circumferencial} shows that the quadrupolar feature of these two waves is well captured on the circumferential profiles, and the results are found to be in agreement with the DNS data for both values of $\sigma$. However, the radial cuts of different instants on Fig.~\ref{fig: DP_Case_C_Radial}, show that better results are obtained for $\sigma=0.9$. This is in accordance with the remarks in Sec.~\ref{sec:covo} where the error grow as the $\sigma$ parameter decreases. In addition, the head of the precursor is more spread out than the reference and thus not well captured by the model. Nevertheless, the second sound, which is generated once the vortex passes through the shock, is correctly predicted for all the characteristic times. Thus it points out that the defect might come from the shock computation, and can be identified as being caused by the non-conservative nature of the energy equation, pointed out in Sec.~\ref{sec: Sod}.

\section{\label{sec:Conclusion}Conclusions}
The present $D2Q9$ HLBM model is an extension to transonic and supersonic flows, both with and without shocks, as the previous model~\cite{Feng2019} has been shown to be restricted to subsonic regimes. Several new elements have been introduced in the present paper, some of which have allowed the extension to supersonic flows.  First, the correction of the trace of the viscous stress tensor, which removes the spurious bulk viscosity. Second, the viscous heat production term in the energy equation has been successfully modeled using LBM combined with finite difference, permitting to enforce the consistency between the two systems and has been validated on smooth flows. Eventually, another discretization scheme for the correction term induced by the closure relation of the lattice was proposed. Through different test cases, these improvements of the original model~\cite{Feng2019} are observed to be not regressive. Most importantly, it permit to dramatically extend the application range of the HLBM model to supersonic flows. This feature demonstrates the importance of the discretization of the correction term and will be investigated in a future work using the von Neumann analysis of the present system~\cite{wissocq2019extended}.

Thus the present HLBM on a low-order, standard lattice has been proven to retrieve the physics of the fully compressible Navier-Stokes-Fourier equation for arbitrary heat capacity ratio and Prandtl number. The use of such a lattice is a higly desirable feature for the implementation of boundary conditions, grid refinement and immersed boundary techniques. It has been shown that it is coherent for low and high Mach number flows with discontinuities, at relatively low viscosity, and only requires to specify the CFL number and a unique numerical parameter $\sigma$. Excellent results were obtained on smooth flow regions. However, in case of discontinuities, the model has been shown to be very dependent on the free parameter $\sigma$. In a future work, this dependency has to be studied thoroughly in order to build a model based on a dynamic $\sigma$, so that a local compromise between stability and accuracy is achieved. Moreover, to enhance the model on shocks at higher pressure ratios, two solutions can be explored: either benching the free parameter on several 1D shock cases in the manner of~\cite{Nie2009}, or using a conservative form of the energy equation. Despite the seemingly straightforward nature of the second solution, it remains a significant challenge up to this day, and to the author's knowledge, no stable conservative form has been found or presented in the HLBM framework.

\appendix

\section{\label{sec:Chapman dev}The Chapman-Enskog development}
Let us start from the corrected discrete velocity Boltzmann BGK equation:
\begin{equation}
\mathcal{B}_i : \frac{D f_i}{D t} = -\frac{f_i-f_i^{eq}}{\tau} + \psi_i ,
\label{eq:DVBE}
\end{equation} 
with $D/D t = \partial/\partial t + c_{i,\alpha} \partial/\partial x_\alpha$ the Lagrangian derivative. 
The Chapman-Enskog development~\cite{chapman1970mathematical} is based on the expansion of the distribution function in terms of the small parameter $\tau$, related to the Knudsen number. Thus using Eq.~(\ref{eq:DVBE}) and following the note described in~\cite{chen2017note}, one can express $f_i$ as:
\begin{equation}
f_i = f_i^{eq} + \tau \psi_i - \tau\frac{D f_i}{D t},
\end{equation}
and further:
\begin{equation}
\begin{split}
f_i &= f_i^{eq} + \tau \psi_i - \tau \frac{D}{D t} \left( f_i^{eq} + \tau \psi_i - \tau \frac{D f_i}{D t} \right) \\
    &= f_i^{eq} + \tau \psi_i - \tau \frac{D}{D t} \left( f_i^{eq} + \tau \psi_i - \tau \frac{D }{D t} \left[ f_i^{eq} + \tau \psi_i - \tau\frac{D f_i}{D t}  \right] \right)\\
    &= ...
\end{split}
\end{equation}
Truncating this expression up to the first-order in $\tau$ leads to:
\begin{equation}
f_i \simeq f_i^{eq}   - \tau \left( -\psi_i + \frac{D}{D t} \left( f_i^{eq}  \right)   \right),
\label{eq:ChapmanO1}
\end{equation}
or more commonly:
\begin{equation}
f_i \simeq f_i^{eq}   + f_i^{1}.
\end{equation}
Here $f_i^{1}$ corresponds to the first-order approximation in $\tau$ of the off-equilibrium part of $f_i$. Although being a first-order approximation, it turns out to be sufficient to recover the macroscopic equations at a Navier-Stokes level~\cite{shan2006kinetic}. Thus the off-equilibrium part of the distribution function $f_i$ is considered to be $f_i^1$.

\bibliographystyle{acm}
% \bibliography{BIBLIO}

\end{document}